\pdfoutput=1

\documentclass[english]{article}

\usepackage[margin=1in]{geometry}
\usepackage[utf8]{inputenc} 
\usepackage[T1]{fontenc}    
\usepackage{hyperref}       
\hypersetup{unicode = true, colorlinks,linkcolor=red,anchorcolor=blue,citecolor=blue,
pdftitle = {On the Stable Resolution Limit of Total Variation Regularization for Spike Deconvolution},
pdfauthor = {Maxime Ferreira Da Costa, Yuejie Chi}}
\usepackage{url}
\usepackage{algorithmic}
\usepackage{bm,bbm}
\usepackage{amsthm,amssymb,amsmath,mathtools}
\usepackage{stmaryrd}
\usepackage{microtype}
\usepackage{xcolor}
\usepackage{graphicx,tabularx,float}
\usepackage{tabularx}
\usepackage{comment}

\makeatletter
\theoremstyle{plain}
\newtheorem{thm}{Theorem}
\newtheorem{defn}[thm]{Definition}
\newtheorem{lem}[thm]{Lemma}
\newtheorem{prop}[thm]{Proposition}

\makeatother

\newcommand\mf[1]{[{\color{blue}MF: #1}]}

\newcommand{\sign}{\mathop{\rm sgn}}
\newcommand{\spann}{\mathop{\rm span}}
\newcommand{\sinc}{\mathop{\rm sinc}}
\newcommand{\diag}{\mathop{\rm diag}}

\newcommand{\argmin}{\mathop{\rm arg\,min}}
\newcommand{\argmax}{\mathop{\rm arg\,max}}

\let\hat\widehat
\let\tilde\widetilde

\newcommand{\pprime}{{\prime\prime}}
\newcommand{\ppprime}{{\prime\prime\prime}}

\title{On the Stable Resolution Limit of Total Variation Regularization for Spike Deconvolution}

\author{Maxime Ferreira Da Costa, Yuejie Chi \\
Department of Electrical and Computer Engineering \\
 Carnegie Mellon University, Pittsburgh, PA 15213, USA\\
Emails: \texttt{\{mferreira,yuejiechi\}@cmu.edu}\footnote{This work is supported in part by Office of Naval Research under the grants N00014-18-1-2142 and N00014-19-1-2404, and in part by National Science Foundation under the CAREER grant ECCS-1650449.}}

\date{October 3, 2019;  \; Revised \today}

\begin{document}

\maketitle

\begin{abstract}
	The stability of spike deconvolution, which aims at recovering point sources from their convolution with a point spread function (PSF), is known to be related to the separation between those sources. When the observations are noisy, it is critical to ensure {\em support stability}, where the deconvolution does not lead to spurious, or oppositely, missing estimates of the point sources. In this paper, we study the resolution limit of stably recovering the support of two closely located point sources using the Beurling-LASSO estimator, which is a convex optimization approach based on  total variation regularization. We establish a sufficient separation criterion between the sources, depending only on the PSF, above which the Beurling-LASSO estimator is guaranteed to return a stable estimate of the point sources, with the same number of estimated elements as that of the ground truth. Our result highlights the impact of PSF on the resolution limit in the noisy setting, which was not evident in previous studies of the noiseless setting. Towards the end, we show that the same resolution limit applies to resolving two close-located sources in conjunction of other well-separated sources.  \\
\end{abstract}

\section{Introduction} \label{sec:intro}

In its classical formulation, the super-resolution problem consists of recovering a stream of localized temporal events, modeled as one-dimensional point sources (or spikes), characterized by their positions and amplitudes, from distorted and noisy observations. This problem finds a myriad of applications in applied and experimental sciences, such as spectrum and modal analysis, radar, sonar, optical imaging, wireless communications and sensing systems. The distortion is often characterized by a shift-invariant point spread function (PSF), acting as a low-pass band-limited filter, on the stream of spikes to recover, in accordance to the physical limitation of the measurement device involved in the acquisition of the point sources \cite{lindberg_mathematical_2012}.

The problem, known as spike deconvolution, comes with a handful of statistical challenges. Of particular interest to this paper is the \emph{support stability} of the reconstruction in the presence of additive noise, defined as the capability of a given estimator to return the exact same number of point sources as that of the ground truth, without spurious or missing elements. This paper studies the support stability of the Beurling-LASSO estimator \cite{candes_towards_2014} to reconstruct two closely located point sources. Despite its apparent simplicity, this setup is of importance both in theory and in practice. In theory, it allows us to develop a deeper insight on the fundamental notion of \emph{resolution limit} -- the minimal distance above which two point sources are said to be distinguishable. In practice, it models the separation of a weak moving target from a strong clutter in radar \cite{chi_range_2009}, and
accurate counting of the number of molecules in super-resolution fluorescence microscopy \cite{lee_counting_2012}.

The Beurling-LASSO estimator is a convex optimization approach with the total variation (TV) regularization, which has been shown to provide exact reconstruction of the point sources in the absence of noise, whenever the point sources are sufficiently separated \cite{candes_towards_2014,fernandez-granda_super-resolution_2016,duval_exact_2015}. The TV regularization, applied to measures, can be regarded as a continuous analog of the standard $\ell_1$ regularization for finite-dimensional vectors, but is advantageous by overcoming the basis mismatch issue \cite{chi_sensitivity_2011}. In this paper, we show that the Beurling-LASSO estimator can also stably reconstruct the support of the two close-located point sources provided that they are separated by a distance that can be computed exactly using a formula depending only on the PSF, revealing the impact of PSF on the stability of spike deconvolution. Our result can be extended to a multi-source setting containing a mixture of two close-located sources and other well-separated sources.

\subsection{Observation model}

We consider a scenario where there are only two point sources to recover. Denoting by $\mathcal{M}(\mathbb{R})$ the set of complex Radon measures over the reals, the signal to resolve is modeled as a measure $\nu_{\star} \in \mathcal{M}(\mathbb{R})$ of the form
\begin{equation}
	\label{eq:signalModel}
	\nu_\star(\tau) = c_1 \delta(\tau - \tau_1) + c_2 \delta(\tau - \tau_2),
\end{equation}
where $\delta(\cdot)$ is the Dirac measure, $\tau_1,\tau_2\in\mathbb{R}$ are the time-domain locations of the two spikes and $c_1,c_2\in\mathbb{C}\backslash\{0\}$ are their non-zero associated complex amplitudes. The continuous-time signal $x(\tau)$ resulted from the convolution of the ground truth measure ${\nu}_{\star}(\tau)$ with the PSF  $g(\tau)$ writes as
\begin{align}
	\label{eq:noiselessObservation}
	x(\tau) & = \left(g \ast \nu_\star \right) (\tau) \nonumber  \\
	     & = c_1 g(\tau-\tau_1) + c_2 g(\tau-\tau_2), \quad \forall \tau\in\mathbb{R},
\end{align}
where $*$ denotes linear convolution. Furthermore, because of the needs of digital processing, one typically takes discrete-time measurements. An idealistic, yet credible approximation of many super-resolution problems encountered in practice is to consider measurements drawn from uniform sampling of the Fourier transform of $x(\tau)$. Let $\mathcal{F}(\cdot)$ be the Fourier transform of a measure in $\mathcal{M}(\mathbb{R})$, defined as
\[
	\mathcal{F}(\mu)(f) = \int_\mathbb{R} e^{-i2\pi f \tau}{\rm{d}} \mu(\tau),\quad \forall \mu \in\mathcal{M}\left( \mathbb{R}\right),\;\forall f \in \mathbb{R},\;\mbox{a.e.}.
\]
The Fourier-domain counterpart of the observation model \eqref{eq:noiselessObservation} becomes
\[
	X(f) = G(f)\cdot\mathcal{F}( {\nu_{\star}})(f), \quad \forall f \in \mathbb{R},\;\mbox{a.e.},
\]
where $X = \mathcal{F}(x)$, $G=\mathcal{F}(g)$ are the Fourier transforms of the signal $x(\tau)$ and the PSF $g(\tau)$, respectively. We assume that the PSF $g(\tau)$ is band-limited, with a bandwidth of $B>0$. Therefore, $G(f)=0$ for every $f$ outside the interval $\left(-\frac{B}{2},\frac{B}{2}\right)$. We further assume an odd number $N=2n+1$ of measurements{\footnote{An odd number of measurements is considered only for clarity and simplification purposes, and does not affect the generality of the results presented in this paper.}} are taken uniformly over the bandwidth $\left(-\frac{B}{2},\frac{B}{2} \right)$. Therefore, the observation vector is given by $\bm{x} = {\{ x_k = X(kB/N) \}}_{k=-n}^n \in \mathbb{C}^N$, corresponding to measuring $X(f)$ at frequencies ${\left\{ kB/N \right\}}_{k=-n}^{n} \subset \left(-\frac{B}{2},\frac{B}{2} \right)$.

For convenience, we introduce a normalized measure $\mu_{\star}\in\mathcal{M}(\mathbb{R})$ as $\mu_{\star}(t) = \frac{N}{B} \nu_\star(N t /B)$ for all $t\in\mathbb{R}$, which by combining with \eqref{eq:signalModel} can be rewritten as,
\begin{equation}
\label{eq:normalizedMeasure}
	\mu_{\star}(t) = c_1 \delta(t - t_1) + c_2 \delta(t - t_2),
\end{equation}
where $t_1 = B \tau_1 / N$ and $t_2 = B \tau_2 / N$ are the normalized  locations of the point sources.
The observations $\bm{x}$ are linked to $\mu_{\star}$ by the linear relation
\begin{align}
	\label{eq:noiselessSamples}
	\bm{x} & = \Phi_{g}(\mu_{\star}).
\end{align}
Here, the measurement operator $\Phi_{g}: \mathcal{M}(\mathbb{R}) \mapsto\mathbb{C}^{N}$ is defined by
\begin{align}
	\label{eq:SamplingOperator}
	\Phi_{g}:        \mu      & \mapsto \diag(\bm{g}) \begin{bmatrix}\mathcal{F}(\mu)(-n),\mathcal{F}(\mu)( -n+1),\dots,\mathcal{F}(\mu)(n)\end{bmatrix}^{\top},
\end{align}
where $\bm{g} = \{ g_k = G(kB/N)\}_{k=-n}^{n} \in\mathbb{C}^N$ is the vector obtained by sampling the Fourier transform of the PSF $g(\tau)$ at frequencies ${\left\{ kB/N \right\}}_{k=-n}^{n}$. Furthermore, notice that the observation operator $\Phi_{g}$ is invariant with respect to integer shifts of the underlying measure $\mu_{\star}$. Thus, one can only hope to identify $\mu_{\star}$ over the set of Radon measure defined over the torus $\mathbb{T}\sim \mathbb{R} \slash \mathbb{Z}$, denoted as $\mathcal{M}(\mathbb{T})$. Without loss of generality, the delays $t_1,t_2$ are normalized within the unit interval, i.e. $t_1,t_2\in [-\frac{1}{2},\frac{1}{2})$\footnote{Since $t_i= B\tau_i /N$, $i=1,2$, and assuming $\tau_i \in [-T/2,T/2)$, where $T$ is the time window of interest, then the ambiguity constraint $t_i \in [-1/2,1/2)$ suggests that the number of measurements should be greater than the time-bandwidth product, i.e. $N \geq T\cdot B$, to avoid aliasing. }.

In the presence of noise or measurement errors, we assume $\bm{x}$ is corrupted by an additive term $\bm{w}$. The observations are given as
\begin{align}
	\label{eq:noisySamples}
	\bm{z} & = \bm{x} + \bm{w}  = \Phi_{g}(\mu_{\star}) + \bm{w},
\end{align}
where ${\|\bm{w} \|}_2 \leq \eta$ is assumed to be bounded for some noise level $\eta>0$.

\subsection{Reconstruction using the total variation minimization framework}

In the absence of noise, the super-resolution problem is defined as recovering $\mu_{\star}$ from the observations $\bm{x}$ and the PSF $g(\tau)$, yielding a linear inverse problem over the set of measures. Clearly, there are many possible measures that lead to the same observations, making the problem ill-posed. It is therefore, necessary to impose structures on the measure of interest, where one of the most widely used structures is a sparsity prior. More precisely, one seeks for the measure $\hat{\mu}_\natural$  with minimal support that is consistent with the observations $\bm{x}$ given in \eqref{eq:noiselessSamples}. Denoting by $\left\Vert\cdot\right\Vert_0$ the ``pseudo-norm'' counting the potentially infinite cardinality of the support of a measure in $\mathcal{M}(\mathbb{T})$, the optimal estimator $\hat{\mu}_\natural$ for the super-resolution problem can be reformulated as the output of the optimization program
\begin{equation}
	\label{eq:optimalEstimator}
	\hat{\mu}_\natural = \argmin_{\mu\in\mathcal{M}(\mathbb{T})} \left\Vert\mu\right\Vert_0\;\; \mbox{s.t.} \;\; \bm{x} =\Phi_{g}(\mu),
\end{equation}
which is known to be unique and equal to the ground truth $\mu_{\star}$ as long as the number of measurements $N$ is at least twice as large as the number of spikes to recover \cite{donoho_optimally_2002}.

However, on the computational front, the estimator \eqref{eq:optimalEstimator} is infeasible due to the combinatorial aspects inherent to the definition of $\left\Vert\cdot\right\Vert_0$. Instead, a convex relaxation of the estimator \eqref{eq:optimalEstimator} is proposed in \cite{candes_towards_2014} to recover the measure. This is achieved by relaxing the cost function by a convex surrogate known as total variation (TV), denoted as $\left\Vert \cdot \right\Vert_{\rm TV}$, whose formal definition will be discussed later. The total variation minimization of measures, equivalent to the atomic norm \cite{chandrasekaran_convex_2012,chi_harnessing_2019}, is a versatile framework that can be adapted to solve a variety of linear inverse problems over continuous dictionaries. The resulting TV estimator, denoted as $\hat{\mu}_0$, is given by
\begin{equation}
	\label{eq:noiselessEstimator}
	\hat{\mu}_0 = \argmin_{\mu\in\mathcal{M}(\mathbb{T})} \; \left\Vert\mu\right\Vert_{\rm TV}\;\; \mbox{s.t.} \;\; \bm{x}=\Phi_{g}(\mu),
\end{equation}
which is a convex program over the set of Radon measures, and can be computed efficiently by solving an associated semidefinite program (see e.g. \cite{bhaskar_atomic_2013}).
In the presence of noisy observations of the form \eqref{eq:noisySamples}, the Beurling-LASSO estimator $\hat{\mu}_{\lambda}$ \cite{de_castro_exact_2012}, also known as the atomic norm denoiser \cite{bhaskar_atomic_2013}, can be used to recover the ground truth. It can be understood as an extension of the celebrated LASSO estimator over the set of measures,  which aims to estimate the ground truth measure by minimizing a sum of the TV norm of the measure and the squared Euclidean norm of the measurement residual \eqref{eq:noiselessSamples}, so that the estimate $\hat{\mu}_{\lambda} $ is written as
\begin{equation}
	\label{eq:noisyEstimator}
	\hat{\mu}_{\lambda} =\hat{\mu}_{\lambda}(\bm{z}) = \argmin_{\mu\in\mathcal{M}(\mathbb{T})} \; \frac{1}{2} \left\Vert \bm{z} - \Phi_{g}(\mu)\right\Vert_2^2 + \lambda \left\Vert \mu \right\Vert_{\rm{TV}},
\end{equation}
where $\lambda>0$ is a regularization parameter drawing a trade-off between the TV norm of the estimate, as well as its fidelity to the observations.

\subsection{Resolution limit of spike deconvolution}

An important question for practical operations is the resolution of spike deconvolution, where one would like to ensure that the reconstructed measure is as close as possible to the ground truth. It has been known for decades that the separation between the spikes,
\begin{equation}\label{eq:separation}
\Delta = \left\vert t_2 - t_1 \right\vert_{\mathbb{T}} \triangleq \inf_{\ell\in\mathbb{Z}} \left\vert t_2 - t_1 + \ell \right\vert,
\end{equation}
which measures the distance over the torus $\mathbb{T}$, plays an important role -- the smaller the separation, the more challenging it is to resolve them. For example, the Rayleigh limit (see e.g. \cite{born_principles_1999}) is a classical \emph{empirical} criterion to characterize the \emph{resolution limit} the super-resolution problem, i.e., the minimal separation between two point sources, above which those sources are said to be distinguishable. In recent years, there has been a resurge of interest in a formal characterization of this limit - both in terms of achievability and impossibility. In particular, stability has been shown to be related to the asymptotic behaviors of the condition number of Vandermonde matrices with nodes on the unit circle \cite{moitra_super-resolution_2015,aubel_theory_2018,batenkov_stability_2018}, which diverges below a critical separation of the spikes. This phase transition induces the existence of a resolution limit under which point sources cannot be resolved in the presence of noise in the asymptotic regime where $N$ tends to infinity \cite{diederichs_well-posedness_2019}, regardless of the algorithm used for reconstruction.

In this paper, we are interested in understanding the robustness of an estimator in the presence of noise. Among the many figures of merit to quantify this robustness, an important criterion is the \emph{support stability} of the estimator, defined below when specialized to the two-spike setting.

\begin{figure}[ht]
	\centering
	\begin{tabular}{cc}
	\includegraphics[width=0.47\textwidth]{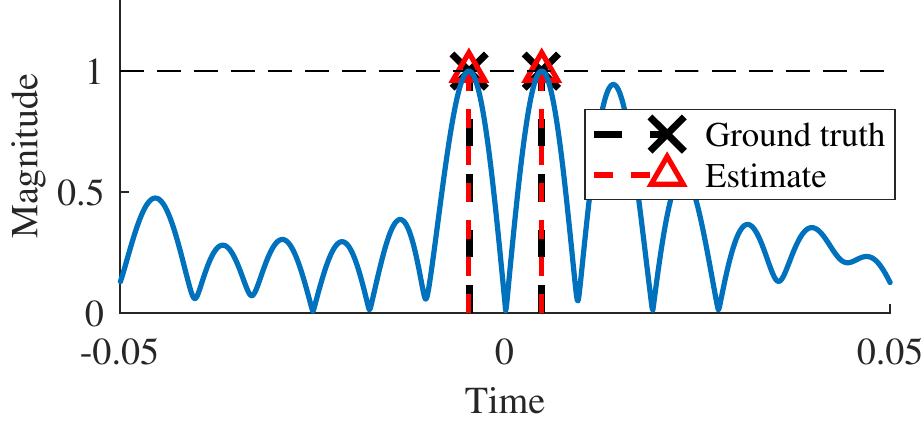}&
	\includegraphics[width=0.47\textwidth]{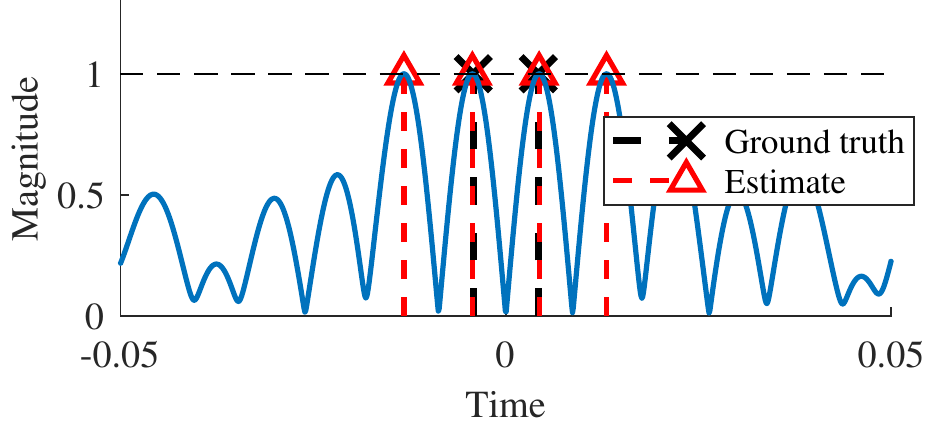} \\
	(a) $\Delta = 1.2/N$ & (b) $\Delta = 1.1/N$
	\end{tabular}
	\caption{Illustration of support stability of the Beurling-LASSO estimator for reconstructing two point sources with different separations $\Delta$ when the PSF is the ideal low-pass filter $g(\tau)=\sinc(\pi \tau)$ and for a number of samples $N=129$ with $\mathrm{SNR}=40$dB. The locations of point sources are estimated as the peaks of the magnitude of the dual polynomial  $\Phi_{g}^{\ast}(\hat{\bm{p}}_{\lambda})$, where $\hat{\bm{p}}_{\lambda}$ is the solution of \eqref{eq:noisyDual}.
	(a): $\Delta = 1.2/N$, the estimator returns exactly two spikes closely located to the ground truth and is support stable. (b): $\Delta = 1.1/N$, the estimator returns two additional spurious spikes and is not support stable.}
	\label{fig:sucessFailureOfBeurlingLASSO}
\end{figure}

\begin{defn}[Support stability] \label{def:support_stability}
 Consider the observations $\bm{z} = \Phi_{g}(\mu_{\star}) + \bm{w}$. An estimator $\hat{\mu}=\hat\mu(\bm{z})$ based on $\bm{z}$ is said to be support stable for a given ground truth measure $\mu_{\star}$ of the form \eqref{eq:normalizedMeasure} if there exists  $\eta > 0$ such that for all $\bm{w}$ with ${\Vert \bm{w} \Vert}_2 < \eta $, the estimate $\hat{\mu}$ is a measure containing two spikes, i.e.
    \[
        \hat{\mu}(\bm{z}) = \hat{c}_1 \delta(t - \hat{t}_1) + \hat{c}_2 \delta(t - \hat{t}_2),
    \]
  and if the estimated parameters satisfy, up to a permutation $\Pi$ of the indices: $\left\vert t_k - \hat{t}_{\Pi(s)} \right\vert_{\mathbb{T}} = \mathcal{O}(\Vert \bm{w} \Vert_2)$ and $\left\vert c_s - \hat{c}_{\Pi(s)} \right\vert = \mathcal{O}(\left\Vert \bm{w} \right\Vert_2)$ for $s=1,2$ in the limit of ${\Vert \bm{w} \Vert}_2 \to 0$.
\end{defn}
This notion, introduced in \cite{duval_exact_2015}, characterizes the capability of an estimator to output a measure containing the exact same number of spikes as that of the ground truth, when the signal-to-noise ratio (SNR) is large enough.  As an example, Fig.~\ref{fig:sucessFailureOfBeurlingLASSO} plots the reconstruction of a ground truth measure containing two spikes using the Beurling-LASSO estimator at $\mathrm{SNR}=40$dB under different separations when the PSF is the ideal low-pass filter. In this illustration, when $\Delta=1.2/N$, the estimator returns exactly two spikes closely located to the ground truth; on the other hand, when $\Delta=1.1/N$, the estimator returns additional spurious spikes that are not consistent with the ground truth, and therefore, is no longer support stable.

\subsection{Overview of the main result}

 \begin{table}[t]
	\centering
	\begin{tabular}{|m{2.9in}|m{1.5in}|m{1.2in}|}
		\hline
		\vspace{0.1in}
		Point spread function \vspace{0.1in}& \vspace{0.1in} \centering Fourier transform \vspace{0.1in} & \centering \vspace{0.1in} $\gamma^\star$ \vspace{0.1in}\tabularnewline
		\hline\hline
		Ideal low-pass: $\sinc(\pi \tau)$                      &   \centering \includegraphics[width=0.212\textwidth]{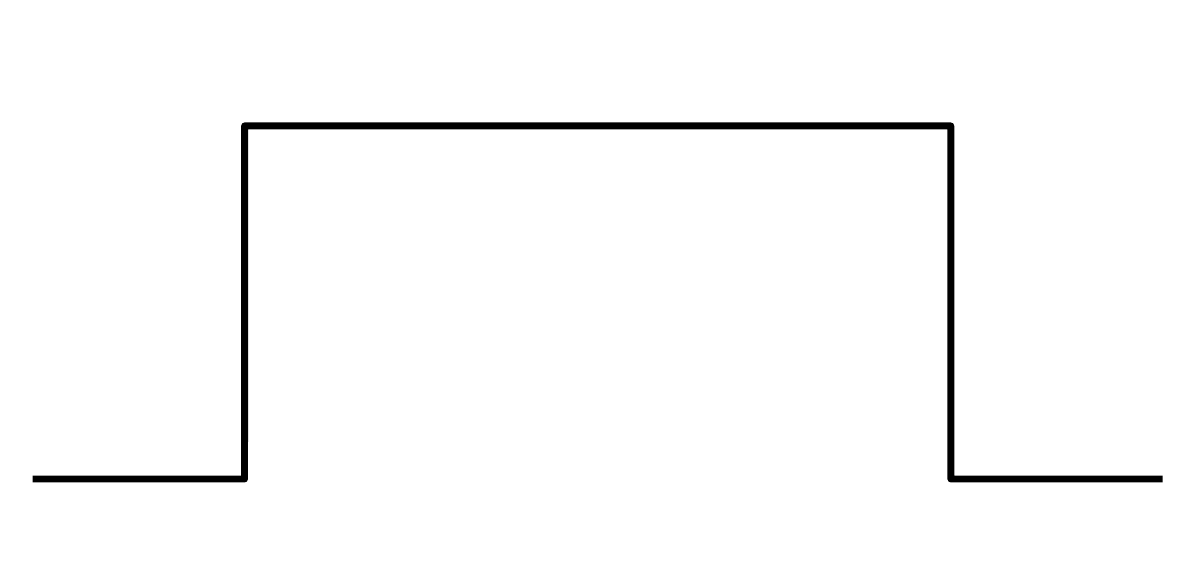}    & \centering 1.132\tabularnewline
		\hline
		Circular low-pass: $J_0(\pi \tau) / \sqrt{\pi \tau} $       & \centering \includegraphics[width=0.212\textwidth]{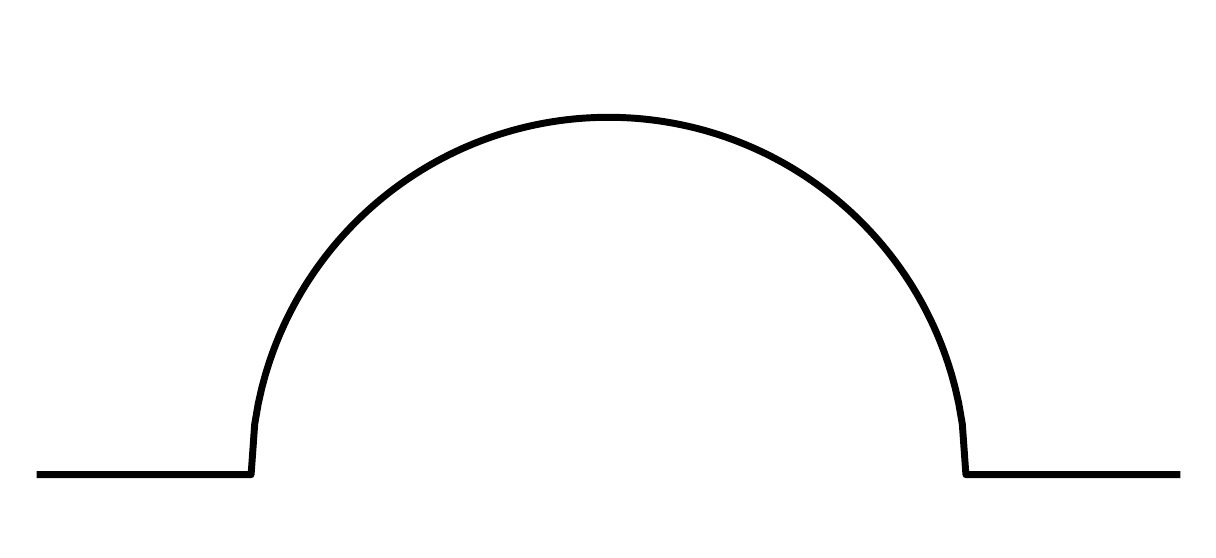}         & \centering 1.253\tabularnewline
		\hline
	    Triangular low-pass: ${\sinc(\pi \tau/2)}^2$                      &   \centering \includegraphics[width=0.212\textwidth]{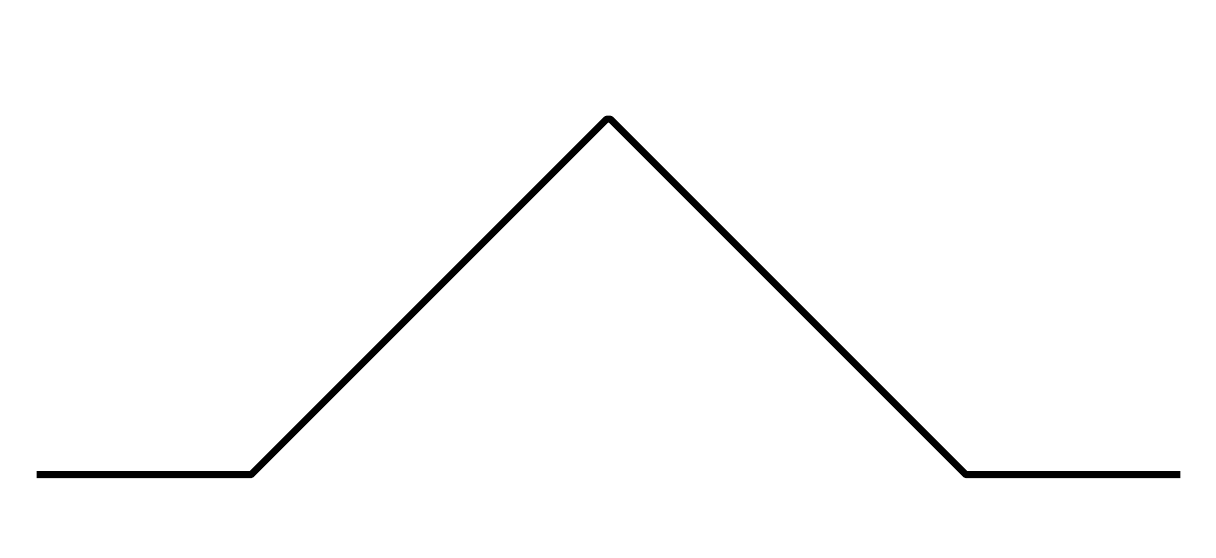}    & \centering 1.449\tabularnewline
		\hline
		Truncated Gaussian: $e^{-\frac{\tau^{2}}{2\sigma^2}}\ast \sinc(\pi \tau)$                  & \centering  \includegraphics[width=0.212\textwidth]{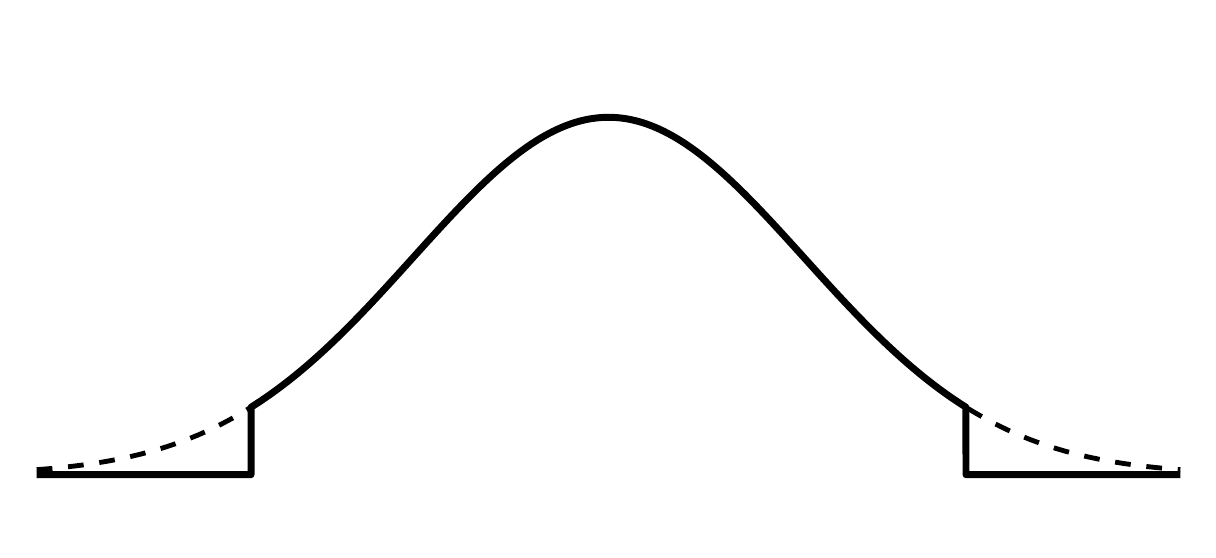}        & \centering (see Fig.~\ref{fig:Resolution} (a))\tabularnewline
		\hline
		Prolate spheroidal wave function: $\psi_{\tau_0}(\tau)$ & \centering \includegraphics[width=0.212\textwidth]{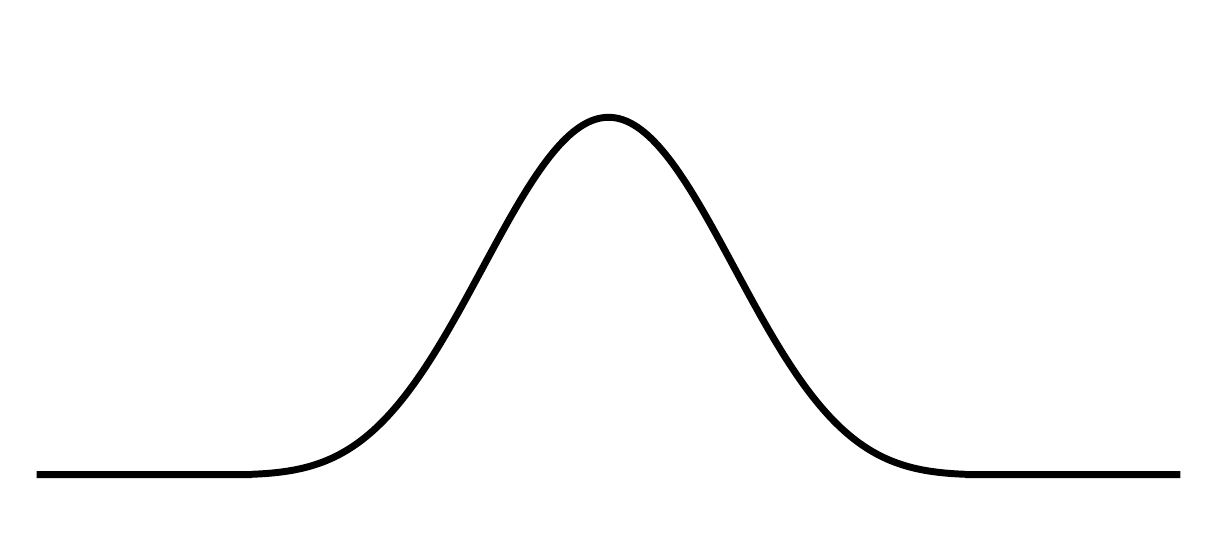}   & \centering (see Fig.~\ref{fig:Resolution} (b))\tabularnewline
		\hline
	\end{tabular}
	\caption{Values of the minimal separation $\gamma^\star$ for commonly encountered point spread functions. Herein, $J_0(\cdot)$ denotes the Bessel function of the first kind, and $\psi_{\tau_0}(\cdot)$ refers to the prolate spheroidal wave function (PSWF) of order 0 for the temporal concentration band $\left[ -\tau_0, \tau_0\right]$ \cite{slepian_prolate_1961,moore_prolate_2004}, i.e.,  the function $g(\cdot)$ with a frequency band $\left(-\frac{1}{2},\frac{1}{2}\right)$ and with $\left\Vert g \right\Vert_{L_2} = 1$ which maximizes the integral
\(
	\int_{-\tau_0}^{\tau_0} {\vert g(\tau) \vert}^2 \mathrm{d}\tau
\)
.}	\label{tab:PSFResolution}
\end{table}

This paper studies the support stability of the Beurling-LASSO estimator \eqref{eq:noisyEstimator} for the reconstruction of a two-spike measure of the form \eqref{eq:signalModel}. We show that, the Beurling-LASSO estimator is support stable,
if the separation $\Delta$ is greater than $\gamma^\star/N$, which can be calculated exactly using the knowledge of the PSF $g(t)$. Our main contribution can be informally summarized in the following statement.

\begin{thm}[main result, informal statement]\label{thm:informal}
	Suppose that the PSF $g(\tau)$ satisfies some mild regularity conditions and is band-limited within $\left(-\frac{B}{2},\frac{B}{2}\right)$. There exists a constant $\gamma^\star$, depending only on $g(\tau)$, such that if a measure $\mu_\star$ of the form \eqref{eq:normalizedMeasure} verifies
	\[
	\Delta:=	\left\vert t_2 - t_1 \right\vert_{\mathbb{T}} > \frac{\gamma^\star}{N},
	\]
	then the Beurling-LASSO estimator is support stable when $N$ is sufficiently large.
\end{thm}

The complete statement of the theorem (c.f. Theorem~\ref{thm:stableResolutionLimit}) provides the formula of the stable resolution limit $\gamma^{\star}$, which can be computed for an arbitrary PSF verifying the hypotheses of the theorem. One highlight of our result is that it links the stable resolution limit directly to the PSF, which is not apparent in the study of the noise-free setting, where the resolution limit of exact recovery is independent of the PSF. This provides a quantitative means to evaluate and compare the choices of different PSFs in imaging and sensing applications. The spectrum of a typical PSF has a decaying shape. A slower decay usually is associated with a smaller resolution limit and better super resolution capabilities for PSFs with the same bandwidth.

\begin{figure}[t]
	\centering
	\begin{tabular}{cc}
	\includegraphics[width=0.47\textwidth]{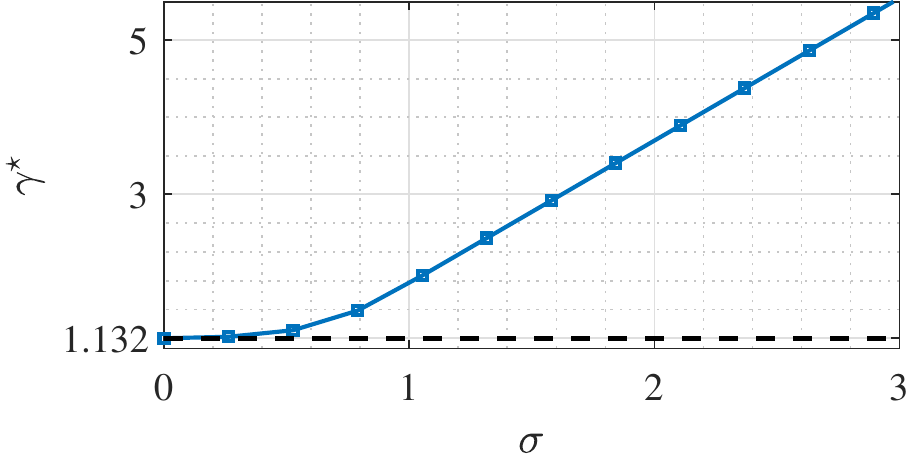} & 	\includegraphics[width=0.47\textwidth]{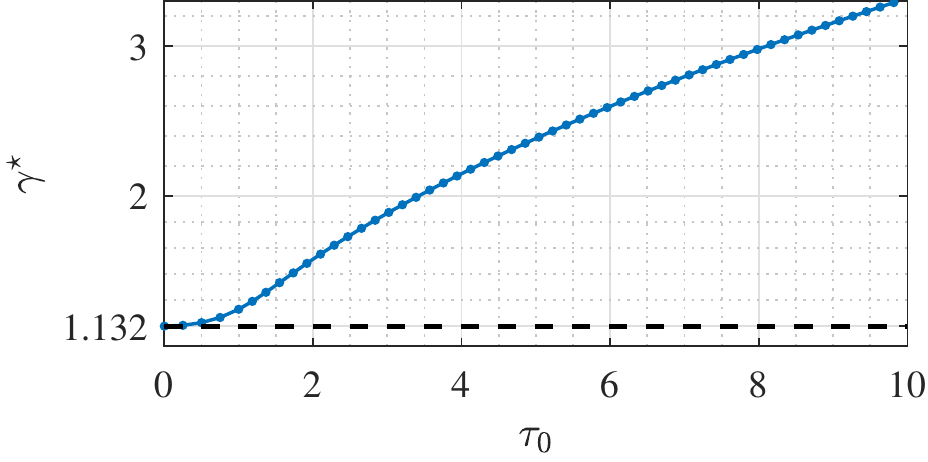}\\
	(a) & (b)
	\end{tabular}
	\caption{The stable resolution limit $\gamma^\star$ for (a) a truncated Gaussian PSF for different values of the parameter $\sigma$, and (b) the prolate spheroidal wave function of order zero $\psi_{\tau_0}$ for different widths of the concentration band $[-\tau_0 , \tau_0]$.}
	\label{fig:Resolution}
\end{figure}

\begin{figure}[ht]
	\centering
	\begin{tabular}{ccc}
	\includegraphics[width=0.31\textwidth]{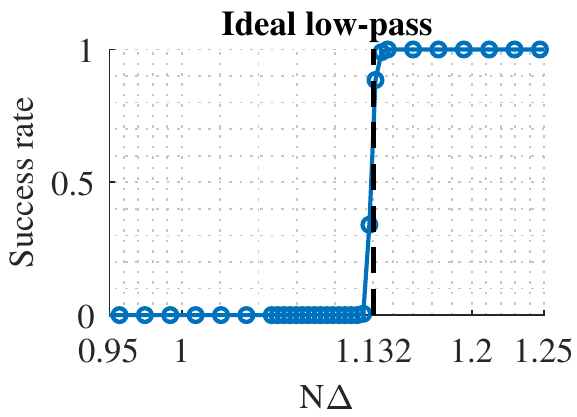}&
	\includegraphics[width=0.31\textwidth]{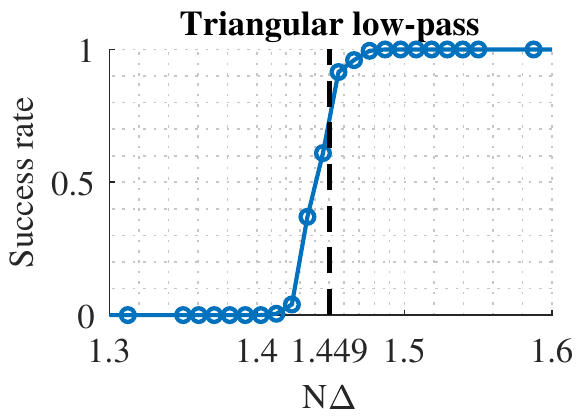}
	&
	\includegraphics[width=0.31\textwidth]{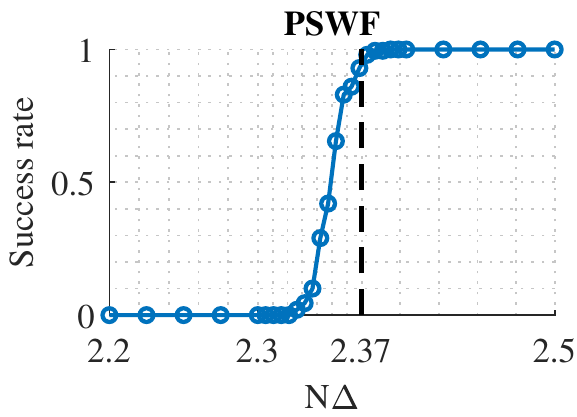}\\
	(a) $g(\tau) = \sinc(\pi \tau)$ & (b) $g(\tau) = {\sinc(\pi \tau/2)}^2$ & (c) $g(\tau) = \psi_{5}(\tau)$
	\end{tabular}
	\caption{Empirical success rates for the Beurling-LASSO estimator to return a measure with two point sources, for three different point spread functions, under additive white Gaussian noise, as a function of the separation parameter $N\Delta$. The support stability threshold $\gamma^{\star}$ predicted by Theorem \ref{thm:informal} is shown in a black dashed line. Here, we set $N=101$, $\mathrm{SNR}=60$dB. The results are averaged over 200 trials.}
	\label{fig:successRate}
\end{figure}

For illustration purposes, Table~\ref{tab:PSFResolution} lists the approximate values of $\gamma_\star$ associated with commonly encountered PSFs, such as ideal low-pass filters, circular low-pass filters, triangular low-pass filters, truncated Gaussian functions, and prolate spheroidal wave functions. Since we focus on the case of only two spikes, the separation condition $\gamma^{\star}$ is smaller than those in \cite{candes_super-resolution_2013,fernandez-granda_super-resolution_2016,ferreira_da_costa_tight_2018} for the noise-free setting, which allows more spikes. In addition, Fig.~\ref{fig:Resolution} illustrates how the stable resolution limit $\gamma^{\star}$ increases while the temporal concentration of the truncated Gaussian function and the prolate spheroidal wave function degenerates. Finally, Fig.~\ref{fig:successRate} compares the constant $\gamma^{\star}$ predicted by Theorem \ref{thm:informal} with the empirical success rates of the Beurling-LASSO estimator for different PSFs, which corroborates the findings of our theory. Our result holds even in the presence of additional point sources, as long as they are well-separated, see Section~\ref{sec:extensions}.

\subsection{Connections to related works}

The resolution limits of spike deconvolution have been studied extensively, including but not limited to \cite{donoho_superresolution_1992,shahram2006statistical,batenkov2019super,demanet2015recoverability,smith2005statistical,liao2016music}. The performance of the TV estimator \eqref{eq:noiselessEstimator} has been studied in the noiseless setting with respect to the separation condition \cite{candes_towards_2014,ferreira_da_costa_tight_2018,duval_exact_2015}. Exact recovery of the TV estimator \eqref{eq:noiselessEstimator} is first guaranteed in \cite{candes_towards_2014}, for an arbitrary number of spikes, given a separation $\Delta > 4/N$ under the proviso that the number of observations $N$ is large enough, which has been later improved to $\Delta \geq 2.56/N$ in \cite{fernandez-granda_demixing_2017}. On the other hand, it is known that TV-regularization can fail whenever $\Delta<2/N$ \cite{ferreira_da_costa_tight_2018}. Furthermore, experimental evidence suggest the existence of a phase transition on the success of \eqref{eq:noiselessEstimator} whenever the minimal separation between any pair of spikes crosses the threshold $\Delta = 2/N$ in the limit of $N$ tending to infinity. With the extra assumption that the number of spikes is exactly two, as in \eqref{eq:normalizedMeasure}, it is shown in \cite{duval_exact_2015} that a separation $\Delta > 1/N$ is necessary to guarantee exact recovery.

The support stability of the Beurling-LASSO estimator is studied in \cite{duval_exact_2015} under the non-degenerate source condition (c.f. Def.~\ref{def:nonDegenerateSourceCondition}), however it is unclear and challenging to establish when this condition will hold for general sources. In \cite{duval_characterization_2017,denoyelle_support_2017}, the support stability of reconstructing positive sources is considered without imposing a minimal separation condition.
 Our main theorem in this paper is achieved essentially via verifying the non-degenerate source condition for the two-spike case with arbitrary signs and the presence of additional well-separated spikes, which is already quite technical and non-trivial.

Furthermore, when the PSF is the ideal low-pass filter, and under additive white Gaussian noise, the stability of TV regularization is studied under various metrics. For example, the stability of an estimator $\hat{\mu}$ in the observation domain is studied in \cite{candes_super-resolution_2013,bhaskar_atomic_2013}, which looks at bounds on $\Vert\Phi_{g}(\hat{\mu} -  \mu_{\star})\Vert_2^2 $. The performance of support detection has been examined in \cite{fernandez-granda_support_2013,tang_near_2015,morgenshtern_super-resolution_2016}, which quantifies the residual of $\hat{\mu}$ outside the support of $\mu_{\star}$, however these guarantees do not ensure the estimate $\hat{\mu}$ contains the same number of spikes as the ground truth $\mu_{\star}$. A trade-off between the separation of the spikes and the error of the parameters is highlighted in~\cite{li_approximate_2018} without resorting to the non-degenerate source condition. However, the required separation for the result in \cite{li_approximate_2018} to hold is quite large and assumes a Gaussian noise.

\subsection{Organization of the paper}
The rest of this paper is organized as follows. Section~\ref{sec:prerequisites} provides some prerequisites on spike deconvolution using the Beurling-LASSO estimator, including background literature. Section~\ref{sec:results} states formally the main theorem of this paper including all technical details. Section~\ref{sec:proofs} proves the main theorem and Section~\ref{sec:extensions} discusses the extension of the main theorem to the reconstruction of multiple point sources. Finally, we conclude in Section~\ref{sec:conclusions}.

\section{Prerequisites} \label{sec:prerequisites}

In this section, we discuss the prerequisites on super resolution using total variation regularization, which are useful to the presentation and analysis of the main result in this paper.

\subsection{Mathematical notations}
The transpose and adjunction of a vector $\bm{v}$ is denoted as $\bm{v}^\top$ and $\bm{v}^\ast$ respectively. The adjoint of a linear operator $\Phi$ is written as $\Phi^\ast$. Vectors of a dimension $N=2n+1$ are indexed between $-n$ and $n$, so that $\bm{v} = [v_{-n},\dots, v_{n}]^{\top}$. For any $t\in\mathbb{T}$, we introduce the atomic vector $\bm{a} \left(t\right)\in\mathbb{C}^N$ and its derivative $\dot{\bm{a}}\left(t\right)\in\mathbb{C}^{N}$~as
\begin{subequations}
	\begin{align*}
		\bm{a}\left(t\right) & =\left[e^{-2\pi i\left(-n\right)t},\dots,e^{-2\pi int}\right]^{\top},\\
		\dot{\bm{a}}\left(t\right) & = \frac{\mathrm{d}\bm{a}(t) }{\mathrm{d}t}= - 2 \pi i\left[-ne^{-2\pi i\left(-n\right)t},\dots,ne^{-2\pi int}\right]^{\top} = -2\pi i \diag(-n ,\cdots, n) \bm{a}(t),
	\end{align*}
\end{subequations}
where $\diag(-n ,\dots, n)$ is a diagonal matrix with diagonal entries $-n ,\cdots, n$. Similarly, we define the $\ell$th order derivative of $\bm{a} \left(t\right)$ as $\bm{a}^{(\ell)}\left(t\right)$. To every vector $\bm{q}\in\mathbb{C}^N$, we associate the trigonometric polynomial $Q(t)$ of degree $n$ such that
\[
	Q(t) = \bm{a}(t)^{\ast}\bm{q} = \sum_{k=-n}^{n} q_k e^{2\pi ikt},\quad \forall t\in\mathbb{T}.
\]
Its derivative satisfies $Q^\prime(t) = \dot{\bm{a}}(t)^{\ast}\bm{q}$ for all $t\in\mathbb{T}$.
The real part and conjugate of a complex number $u$ is denoted as $\Re(u)$ and $\bar{u}$, and the sign of a non-zero complex number is given by $\sign(u) = u/|u|$. For any two vectors $\bm{z},\bm{p}\in\mathbb{C}^N$, we denote by $\left\langle\cdot,\cdot\right\rangle_\mathbb{R}$ the real inner product $\left\langle \bm{z}, \bm{p}\right\rangle_\mathbb{R} = \Re\{\bm{z}^\ast \bm{p}\}$, and we denote by $\bm{z} \odot \bm{p}\in\mathbb{C}^N$ their element-wise product.

The vector space of continuous functions from $\mathbb{T}$ to $\mathbb{C}$,  denoted as $\mathcal{C}(\mathbb{T})$, is endowed with the supremum norm $\Vert \cdot \Vert_\infty$. The total variation norm $\Vert \cdot \Vert_{\rm TV}$, defined as the dual norm of $\Vert \cdot \Vert_\infty$,  is given as
\[
	\forall \mu\in\mathcal{M}(\mathbb{T}),\quad \Vert \mu \Vert_{\rm TV}=\sup_{\substack{h\in\mathcal{C(\mathbb{T})}\\
			\left\Vert h\right\Vert _{\infty}\leq 1
		}
	}\Re\left[\int_\mathbb{T}\overline{h\left(t\right)}\mathrm{d}\mu\left(t\right)\right].\label{eq:TVNorm-Definition}
\]

Given two functions $g$ and $h$ that depend on $N$, we use the classical Landau's notations $g = \mathcal{O}(h)$, $g=o(h)$ and $g=\omega(h)$ to denote $\lim_{N\to \infty} \left\vert \frac{g(N)}{h(N)} \right\vert = C$ for some $C\in\mathbb{R}$, $\lim_{N\to \infty} \left\vert \frac{g(N)}{h(N)}\right\vert = 0 $, and $\lim_{N\to \infty} \left\vert \frac{g(N)}{h(N)} \right\vert = \infty$, respectively.

\subsection{Tightness of total variation minimization}

In the noiseless setting, the TV estimator \eqref{eq:noiselessEstimator} is said to be tight if its output $\hat{\mu}_0$ is equal to the output $\hat{\mu}_\natural$ of the estimator \eqref{eq:optimalEstimator}. As for many other convex optimization-based methods for solving inverse problems, the Lagrangian duality theory can be leveraged to derive tightness guarantees. The Lagrange dual problem associated to \eqref{eq:noiselessEstimator} reads
\begin{align}
	\mathcal{D}_0\left(\bm{x} \right) = \argmax_{\bm{p}\in\mathbb{C}^{N}} & \left\langle \bm{x},\bm{p}\right\rangle _{\mathbb{R}} \nonumber              \\
	\mathrm{subject\,to\,}                                                     &\, \Vert \Phi_{g}^\ast(\bm{p}) \Vert_{\infty} \leq 1,\label{eq:noiselessDual}
\end{align}
where the adjoint of the operator $\Phi_{g}$ in \eqref{eq:SamplingOperator} is given by
	\begin{align}
		\label{eq:AdjointSamplingOperator}
		\Phi_{g}^\ast:\;\mathbb{C}^{N} & \to \mathcal{M}(\mathbb{T}) \nonumber                                                                                                                            \\
		\bm{p}                           & \mapsto \Phi_{g}^\ast(\bm{p})(t) = \sum_{k=-n}^{n} p_k \bar{g}_{k} e^{i 2\pi k t}, \nonumber                                               \\
																		 & \phantom{\mapsto \Phi_{g}^\ast(\bm{p})(\tau)} \;= \bm{a}\left(t\right)^{\ast}\diag(\overline{\bm{g}})\bm{p},\quad\forall t\in \mathbb{T}.
	\end{align}
In other words, $\Phi_{g}^\ast$ associates any $\bm{p}\in\mathbb{C}^N$ with a trigonometric polynomial $Q(t)= \bm{a}\left(t\right)^{\ast} \bm{q}$ of degree at most $n$, where $\bm{q} = \diag(\overline{\bm{g}}) \bm{p}$. Moreover, as the restriction of the feasible set of \eqref{eq:noiselessDual} to the span of the operator $\Phi_{g}$ is compact, the set of solutions $\mathcal{D}_0\left(\bm{x}\right)$ is non-empty as long as $\bm{x}$ is a consistent observation under the observation model \eqref{eq:noiselessSamples}.

It is now well understood that the tightness of TV regularization is characterized by the existence of a so-called \emph{dual certificate} \cite{candes_towards_2014}: a function lying in the feasible set of the dual program \eqref{eq:noiselessDual}, and satisfying certain extremal interpolation properties. Considering an input measure with only two point sources of the form \eqref{eq:signalModel}, the corresponding dual certificate is defined as follows.

\begin{defn}[Dual certificate \cite{candes_towards_2014}]
	\label{def:dualCertificate}
	A vector $\bm{p}\in\mathbb{C}^N$ is said to be a dual certificate for the optimization problem \eqref{eq:noiselessEstimator} with an input $\mu_{\star}$ of the form \eqref{eq:normalizedMeasure} if and only if the trigonometric polynomial $Q(t) = \bm{a}\left(t\right)^{\ast} \bm{q}$ with a coefficient vector $\bm{q} = \diag(\overline{\bm{g}}) \bm{p}$ verifies the conditions
	\begin{subequations}
	\label{eq:dualCertificateConditions}
		\begin{align}
			\label{eq:dualCertificateConditions-interp1}Q(t_1)  & = \sign(c_1)       ,                                           \\
			\label{eq:dualCertificateConditions-interp2}Q(t_2)  & = \sign(c_2)   ,                                               \\
			\label{eq:dualCertificateConditions-extremalConstraint}\vert Q(t) \vert & < 1,\qquad \forall t\in\mathbb{T}\backslash\{t_1,t_2\}.
		\end{align}
	\end{subequations}\end{defn}
It can easily be verified that any dual certificate $\bm{p}$ achieves dual optimality with a dual objective $\left\langle \bm{x} ,\bm{p}\right\rangle _{\mathbb{R}} = \Vert \mu_{\star} \Vert_{\rm{TV}}$. In fact, by a duality argument, if such a certificate exists, $\mu_{\star}$ is the unique solution of~\eqref{eq:noiselessEstimator}. As a result, showing the existence of a dual certificate of a given instance of the total variation program~\eqref{eq:noiselessEstimator} provides a constructive approach to prove the tightness of the TV estimator $\hat{\mu}_0$. We refer the reader to~\cite{candes_towards_2014, bendory_robust_2016} for further discussion of this property.

\subsection{Stability of the Beurling-LASSO estimator}\label{subsec:StabilityBeurlingLASSO}

Moving to the noisy case, where we aim to recover $\mu_{\star}$ from the noisy observations \eqref{eq:noisySamples}, it becomes necessary to invoke the Beurling-LASSO estimator \eqref{eq:noisyEstimator}. Due to the existence of noise, it is no longer possible to recover $\mu_{\star}$ perfectly. However, we hope the estimator is stable, so that the estimate $\hat{\mu}_\lambda$ is close to the ground truth measure $\mu_{\star}$ when the noise $\bm{w}$ and the regularization parameter $\lambda$ are small enough. More precisely, we are interested in the support stability as defined in Def.~\ref{def:support_stability}, which is a quite strong metric carrying the desirable notion of maintaining a faithful estimate of each individual spike without incurring spurious or missing spikes. The \emph{support stability} of the Beurling-LASSO estimator $\hat{\mu}_\lambda$ in \eqref{eq:noisyEstimator} has been studied for a broad class of measurement operators in \cite{duval_exact_2015}. The results, again, are derived from an analysis of the Lagrange dual problem of the estimator $\hat{\mu}_\lambda$, which is given as
\begin{align}
	\hat{\bm{p}}_{\lambda}= \hat{\bm{p}}_{\lambda}(\bm{z})=\argmax_{\bm{p}\in\mathbb{C}^{N}} & \left\langle \bm{z},\bm{p}\right\rangle _{\mathbb{R}}-\frac{\lambda}{2}\left\Vert \bm{p}\right\Vert _{2}^{2} \nonumber \\
	\mathrm{subject\,to\,}                                            & \Vert \Phi_{g}^\ast(\bm{p}) \Vert_{\infty} \leq 1,\label{eq:noisyDual}
\end{align}
Note that, contrary to \eqref{eq:noiselessDual}, the solution $\hat{\bm{p}}_\lambda $ is unique for every $\bm{z}$ and $\lambda>0$, as \eqref{eq:noisyDual} can equivalently be interpreted as the projection of $\lambda^{-1}\bm{z}$ onto the convex feasible set. As the regularization parameter $\lambda$ tends to $0$, the output $\hat{\bm{p}}_{\lambda}(\bm{x})$ of the dual problem \eqref{eq:noisyDual} applied on the ground truth observations $\bm{x} = \Phi_{g}(\mu)$  converges towards the element $\hat{\bm{p}}_{\min}$ of the solution set $\mathcal{D}_0\left(\bm{x}\right)$ with the minimal norm so that \cite{duval_exact_2015}
\begin{equation}
	\label{eq:minimalNormCertificate}
	 \lim_{\lambda \to 0} \hat{\bm{p}}_\lambda(\bm{x}) = \argmin \left\{ \left\Vert\bm{p} \right\Vert_2\,:\,{\bm{p}\in\mathcal{D}_0(\bm{x})} \right\}: = \hat{\bm{p}}_{\min}.
\end{equation}
The minimal norm solution $\hat{\bm{p}}_{\min}$ to the dual of the TV estimator \eqref{eq:noiselessEstimator} therefore encodes the behavior of the Beurling-LASSO estimator when both the noise level $\Vert \bm{w} \Vert_2$ and $\lambda$ tend to $0$. In fact, the support stability of Beurling-LASSO can be guaranteed for a measure $\mu_{\star}$ if some algebraic properties on $\hat{\bm{p}}_{\min}$ can be verified. These properties introduced in \cite{duval_exact_2015} as \emph{non-degenerate source conditions} are recalled in the following definition.

\begin{defn}[Non-degenerate source condition \cite{duval_exact_2015}]
\label{def:nonDegenerateSourceCondition}
    A measure $\mu_{\star}$ of the form \eqref{eq:normalizedMeasure} is said to verify the non-degenerate source condition with respect to the measurement operator $\Phi_{g}$ if and only if the trigonometric polynomial $\hat{Q}_{\min}(t) = \bm{a}\left(t\right)^{\ast}\hat{\bm{q}}_{\min}$ with coefficients $\hat{\bm{q}}_{\min} = \diag(\bar{\bm{g}})\hat{\bm{p}}_{\min}$, where $\hat{\bm{p}}_{\min}$ is defined in \eqref{eq:minimalNormCertificate}:
    \begin{enumerate}
        \item verifies the dual certificate conditions in \eqref{eq:dualCertificateConditions},
        \item has non-vanishing second derivatives at the source locations, or equivalently
    		      \begin{align}
			      \label{eq:nonDegenerateSourceCondition}
			      \frac{{\rm d}^{2}}{{\rm d}t^{2}}\left|\hat{Q}_{\min} \left(t_s\right) \right| & <0, \quad s = 1,2.
		      \end{align}
    \end{enumerate}
\end{defn}

The first condition in the above definition essentially requires that the minimal norm solution $\hat{\bm{p}}_{\min}$ is a valid dual certificate for the TV estimator \eqref{eq:noiselessEstimator} with noiseless input $\bm{x}$ and, therefore, guarantees its tightness. In view of \eqref{eq:minimalNormCertificate}, this ensures that the low-noise limit of the Beurling-LASSO estimator \eqref{eq:noisyEstimator} can recover the ground truth measure $\mu_{\star}$, which is a natural requirement for support stability. The second condition adds an additional constraint, which enforces the certificate to have a strictly concave modulus around the location of the point sources, which, roughly speaking, ensures no spurious spikes will be introduced when adding a little bit of noise. Together, these conditions are used to guarantee the support stability of the Beurling-LASSO estimator in the low-noise regime in \cite{duval_exact_2015}, that is recalled in the following proposition.
\begin{prop}[Support stability of Beurling-LASSO \cite{duval_exact_2015}]
	\label{prop:supportStability}
	Suppose that the ground truth measure $\mu_{\star}$ verifies
	the non-degenerate source condition with respect to the sampling operator $\Phi_{g}$ in Def.~\ref{def:nonDegenerateSourceCondition}. Then there exists $\alpha > 0$ such that the Beurling-LASSO estimator $\hat{\mu}_\lambda(\bm{z})$ applied to the noisy measurements $\bm{z}=  \Phi_{g}\left(\mu_{\star}\right) + \bm{w} $ with a regularization parameter $\lambda = \alpha^{-1}\left\Vert \bm{w} \right\Vert_2$  is support stable in the sense of Def.~\ref{def:support_stability}.
\end{prop}

Proposition \ref{prop:supportStability} suggests a constructive approach to prove the support stability of the Beurling-LASSO estimator, namely, by verifying the minimal norm solution $\hat{\bm{p}}_{\min}$ associated with a measure $\mu_{\star}$ satisfies the non-degenerate source condition. However, the original work \cite{duval_exact_2015} does not provide explicit means to verify this condition. Subsequent works \cite{duval_characterization_2017,denoyelle_support_2017} studied the special case of positive sources. Nevertheless, it remains unclear when these conditions are verified for general sources and PSFs. The main theorem, presented in the next section, is built upon verifying the non-degenerate source condition for a two-spike measure with arbitrary coefficients and PSFs satisfying some mild regularity conditions.

\section{Main Result} \label{sec:results}

This section formally introduces the main contribution of this paper, which is to provide a sufficient separation condition between the two spikes of a measure of the form \eqref{eq:normalizedMeasure} to guarantee the support stability. The provided bound depends only on the PSF $g$, and more specifically on its auto-correlation function and successive derivatives. The presented result is achieved via verifying the non-degenerate source condition presented in Def.~\ref{def:nonDegenerateSourceCondition}, whose proof will be detailed in Section~\ref{sec:proofs}.

We denote by $\kappa = \mathcal{K}(g) \in L_2$ the auto-correlation of the PSF $g\in L_2$,
defined as
\begin{equation} \label{eq:autocorrelation}
	\mathcal{K}\left(g\right)\left(\tau \right) = \int_\mathbb{R} \overline{g(u)}g(\tau +u){\rm{d}}u,\quad \forall \tau\in\mathbb{R}.
\end{equation}
Denote two auxiliary functions $u_{\beta},v_{\beta} \in L_2$ defined for every $\beta > 0$ as
\begin{subequations}
\label{eq:uAndvDefinition}
\begin{align}
		u_{\beta}(\tau) &= \kappa\left(\tau-\frac{\beta}{2}\right) +  \kappa\left(\tau+\frac{\beta}{2}\right),\\
v_{\beta}(\tau) &= \kappa\left(\tau-\frac{\beta}{2}\right) -  \kappa\left(\tau+\frac{\beta}{2}\right).
\end{align}
\end{subequations}
The above two functions describe the auto-correlations of a signal produced by two point sources separated by a distance $\beta$ with same and opposite signs, respectively. When $g$ is real, the $\kappa$ is even, and we point out that the functions $u_{\beta},v_{\beta}$ are even and odd, respectively. We are now ready to state the main theorem of this paper.
\begin{thm}[Stable resolution limit of Beurling-LASSO]
	\label{thm:stableResolutionLimit} Suppose that the PSF $g$ satisfies the following regularity conditions (H1)-(H4).
	\begin{enumerate}
	    \item[(H1)] $g\in L_2$ is real and non-zero, \emph{i.e.} $\exists \tau\in\mathbb{R},\;g(\tau) \neq 0$.
	    \item[(H2)] $G=\mathcal{F}\left(g\right)\in L_2$ is band-limited within $B$, \emph{i.e.}
	    $G(f) = 0$, $ \forall \left\vert f \right\vert > B/2$.
	    \item[(H3)] $G$ is bounded, \emph{i.e.} $\sup_{f\in\mathbb{R}} \left\vert G(f) \right\vert = \sup_{f\in [-B/2,B/2]} \left\vert G(f) \right\vert < \infty$.
	    \item[(H4)] For $\kappa = \mathcal{K}(g) \in L_2$ and its first three derivatives, $\kappa^{(\ell)}$ for $\ell = 0,1,2,3$,
	    \begin{subequations}
	       	    \begin{align}
	    \forall N\in 2\mathbb{Z} + 1, \quad S_{\ell}(N) &\triangleq \sup_{t\in\left[-\frac{1}{2},\frac{1}{2}\right]} \left\vert \sum_{\substack{k\in\mathbb{Z} \\ k \neq 0}} \kappa^{(\ell)} \left(\frac{N}{B}\left(t+k\right)\right) \right\vert < \infty , \\
	    \lim_{N\to \infty} S_{\ell}(N) &= 0.
	    \end{align}
	    \end{subequations}
	\end{enumerate}
Let $\gamma^\star$, depending only on the PSF $g$, be defined as $\gamma^\star = \max\{\gamma^\star_1,\gamma^\star_2,\gamma^\star_3\}>0$  with
	\begin{subequations}
	    \label{eq:gamma_star}
		\begin{align}
		    \gamma^\star_1 &= B \sup_{\beta > 0}\left\{ \sup_{\tau \geq 0} \left\vert \tilde{s}_{\beta}\left(\tau\right) \right\vert > \tilde{s}_{\beta}\left(\frac{\beta}{2}\right) \right\}, \label{eq:gamma_star_1}\\
		    \gamma^\star_2 &= B \sup_{\beta > 0}\left\{ \sup_{\tau \geq 0} \left\vert \tilde{r}_{\beta}\left(\tau\right) \right\vert > \tilde{r}_{\beta}\left(\frac{\beta}{2}\right) \right\}, \label{eq:gamma_star_2}\\
			\gamma^\star_3 &= B \sup_{\beta > 0} \left\{-{{\kappa^{\pprime}}(0)}^2 + {{\kappa^{\pprime}}(\beta)}^2 - \kappa^{\prime}(\beta)\kappa^{\ppprime}(\beta) \geq 0\right\}, \label{eq:gamma_star_3}
		\end{align}
	\end{subequations}
where the intermediate functions $\tilde{s}_{\beta}\left(\tau\right)$,  $\tilde{r}_{\beta}\left(\tau\right)$ are further defined, for any $\beta > 0$ and $\tau \in \mathbb{R}$ as
\begin{subequations}
    \label{eq:rtildeAndStilde}
	\begin{align}
		    \tilde{s}_{\beta}\left(\tau\right) &= \left( -\kappa^{\pprime}\left(0\right) - \kappa^{\pprime}\left(\beta\right) \right)v_{\beta}\left(\tau\right) - \kappa^{\prime}\left(\beta\right)u^{\prime}_{\beta}\left(\tau\right),\\
	    \tilde{r}_{\beta}\left(\tau\right) &= \left( -\kappa^{\pprime}\left(0\right) + \kappa^{\pprime}\left(\beta\right) \right)u_{\beta}\left(\tau\right) + \kappa^{\prime}\left(\beta\right)v^{\prime}_{\beta}\left(\tau\right).
	\end{align}
\end{subequations}
Then there exists $N_0\in\mathbb{N}$ such that, for every $N\geq N_0$ and every $\mu_{\star}$ of the form \eqref{eq:normalizedMeasure} with
\begin{equation}
    \label{eq:thm-separationCondition}
  \Delta:=  \left\vert t_1 - t_2 \right\vert_{\mathbb{T}} > \frac{\gamma^\star}{N},
\end{equation}
there exists $\alpha > 0$ such that the Beurling-LASSO estimator $\hat{\mu}_\lambda\left(\Phi_{g}\left(\mu_\star\right) + \bm{w} \right)$ with the regularization parameter $\lambda = \alpha^{-1}\left\Vert \bm{w} \right\Vert_2$  is support stable.
\end{thm}

Theorem~\ref{thm:stableResolutionLimit} provides an explicit means to compute $\gamma^{\star}$, based on the evaluation of \eqref{eq:gamma_star}, for a given PSF satisfying the regularity conditions. The key quantities, $\gamma^\star_k$, $k=1,2,3$, are suprema of continuous functions where the complexity of the computation essentially depends on the variations and smoothness of $\kappa$. It is worth noticing that these quantities are also independent of the bandwidth $B$ through a re-scaling the PSF via a transform $g\left(\tau\right) \gets g\left(c\tau\right)$ for some $c > 0$, and can therefore be computed for a PSF with unit bandwidth $B = 1$. Table~\ref{tab:PSFResolution} provides several examples of stable resolution limits for PSFs frequently encountered in practice. The separation condition \eqref{eq:thm-separationCondition} can be equivalently interpreted in terms of the  delays $\tau_1,\tau_2$ of the unnormalized measure $\nu_\star$ as in \eqref{eq:signalModel} as
\begin{equation}
B \left\vert \tau_2 - \tau_1 \right\vert > \gamma^{\star},
\end{equation}
provided the spikes to be localized are in the interval $\left( -N/2B, N/2B \right)$.
In addition, our results allow arbitrary coefficients of the spikes, as long as they are sufficiently separated above the resolution limit.

If we impose a strictly stronger assumption that the PSF decays ``reasonably fast'', i.e, there exist constants $C_{\ell} > 0$, $\ell = 0,1,2,3$ and $\delta > 0$ such that
\begin{equation}
\label{eq:thm-simplerAssumption}
    \left\vert g^{(\ell)}(\tau) \right\vert \leq \frac{C_{\ell}}{1+\left\vert \tau \right\vert^{1+\delta}},\quad \forall\tau\in\mathbb{R},\;\ell = 0,1,2,3,
\end{equation}
then Assumption is automatically verified, which greatly eases the applicability of Theorem \ref{thm:stableResolutionLimit}. Although \eqref{eq:thm-simplerAssumption} can be verified for the majority of the PSFs encountered in practical applications, it excludes some PSFs of theoretical interest, such as the ideal low-pass function $g(\tau) = \mbox{sinc}\left(\pi \tau \right)$, and the truncated Gaussian kernel.

\section{Proof of Theorem \ref{thm:stableResolutionLimit}} \label{sec:proofs}

This section aims to prove Theorem \ref{thm:stableResolutionLimit}. Recall Proposition \ref{prop:supportStability}, which provides a sufficient condition to establish the support stability of the Beurling-LASSO estimator \eqref{eq:noisyEstimator}. The proof essentially consists of deriving a sufficiently large constant $\gamma^\star$ above which a ground truth measure $\mu_\star$ verifies the non-degenerate source condition (c.f. Definition \ref{def:nonDegenerateSourceCondition}) outlined in Proposition \ref{prop:supportStability} when $N$ is large enough. The proof is divided into four major steps.
\begin{enumerate}
	\item Using invariance properties of the non-degenerate source condition, we start by reducing the problem to reconstructing  a ``canonical'' measure $\mu_{\star}$ exhibiting useful symmetry and simplifying the ensuing calculations.
	\item Next, we introduce a so-called minimal vanishing derivative polynomial $Q_V(t) = \bm{a}(t)^{*}\diag\left(\bar{\bm{g}}\right)\bm{p}_V$, where $\bm{p}_V$ is the solution to a simple linear system depending on the parameters of the canonical measure $\mu_{\star}$. Leveraging Lemma \ref{lem:equivalence} \cite{duval_exact_2015}, we show that it is enough to show $Q_V(t)$ satisfies \eqref{eq:dualCertificateConditions} and \eqref{eq:nonDegenerateSourceCondition}.
	\item We next study the limiting behavior of $Q_V(t)$, where it converges towards a band-limited function $\mathcal{Q}_V(Nt/B)$ as $N\to\infty$. Furthermore, we show that, if $\mathcal{Q}_V(\tau)$ verifies the conditions in \eqref{eq:conditionOnLimitQ}, then $Q_V(t)$ satisfies \eqref{eq:dualCertificateConditions} and \eqref{eq:nonDegenerateSourceCondition}, provided that $N$ is large enough.
	\item Finally, we derive a sufficient separation condition $\gamma_{\star}$ above which the limit function $\mathcal{Q}_V(t)$ verifies the conditions \eqref{eq:conditionOnLimitQ} whenever $N\Delta > \gamma_\star$. We conclude on the statement of Theorem \ref{thm:stableResolutionLimit} by applying Proposition~\ref{prop:supportStability}.
\end{enumerate}

\subsection{Canonical reduction of the problem}
We start by reducing the problem to a simpler form without loss of generality by exploiting the following invariances of the non-degenerate source condition.

\begin{lem}[Invariances of the non-degenerate source condition]
\label{lem:invarianceOfNonDegenerateSourceCondition}
Suppose that a measure $\mu_1\in\mathcal{M}(\mathbb{T})$ of the form \eqref{eq:normalizedMeasure} verifies the non-degenerate source condition given in Def. \ref{def:nonDegenerateSourceCondition} with respect to the sampling operator $\Phi_g$. The follow statements hold:
\begin{itemize}
    \item \emph{Translation invariance:} for any $t_0 \in \mathbb{T}$, the measure $\mu_2(t) = \mu_1(t-t_0)$ for all $t\in\mathbb{T}$ verifies the non-degenerate source condition for the same operator.
    \item \emph{Scaling invariance:} for any $c>0$ and $\theta\in [0,2\pi)$, the measure $\mu_2(t) = ce^{i\theta}\mu_1(t)$ for all $t\in\mathbb{T}$ verifies the non-degenerate source condition for the same operator.
    \item \emph{Time reversal invariance:} the measure $\mu_2(t) = \mu_1(-t)$ for all $t\in\mathbb{T}$ verifies the non-degenerate source condition for the same operator.
\end{itemize}
\end{lem}

A proof of Lemma \ref{lem:invarianceOfNonDegenerateSourceCondition} is provided in Appendix \ref{sec:ProofOfInvarianceOfNonDegenerateSourceCondition}. Leveraging this result, we conclude that the non-degenerate source conditions depends only on the separation $\Delta \triangleq   \left\vert t_2 - t_1 \right\vert_{\mathbb{T}}$ between the two point sources and the angle $\theta\triangleq\arg(c_2/c_1)\;\mbox{mod}\,\pi$ between their complex amplitude. For any $\Delta \in (0,\frac{1}{2}]$ and $\theta \in [0,\pi]$, we define the \emph{canonical} measure $\mu_{\star}(\Delta,\theta)$ as
\begin{equation}
	\label{eq:canonicalMeasure}
	\mu_{\star}(\Delta,\theta) = e^{-i\frac{\theta}{2}}\delta\left(t+\frac{\Delta}{2} \right) + e^{i\frac{\theta}{2}}\delta\left(t-\frac{\Delta}{2}\right).
\end{equation}
We can now restrict our analysis to canonical measures of the form $\mu_\star = \mu_{\star}(\Delta,\theta)$ without loosing any generality, and exploit the Hermitian symmetry of $\mu_\star$ to simplify the ensuing calculations.

\subsection{The minimal vanishing derivative polynomial}\label{subsec:Minimal-vanishing-pre-certificate}

The minimal norm solution $\hat{\bm{p}}_{\min}$ (defined in \eqref{eq:minimalNormCertificate}) associated to the measure $\mu_{\star} = \mu_{\star}(\Delta,\theta)$  can be equivalently interpreted as a projection onto the spectrahedra of bounded trigonometric polynomials. Such projection is difficult to derive analytically, limiting the ability to establish the non-degenerate source condition (c.f. Def.~\ref{def:nonDegenerateSourceCondition}) through a direct analysis of $\hat{\bm{p}}_{\min}$. To bypass this problem, it is proposed in \cite{duval_exact_2015} to study instead the behaviors of a surrogate vector $\bm{p}_{V}$, defined in the present context as the unique solution to the following quadratic problem:
\begin{align}
	\label{eq:minimalVanishingPreCertificate}
	\bm{p}_V = & \argmin_{\bm{p}\in\mathbb{C}^{N}}\left\Vert \bm{p}\right\Vert _{2} \nonumber\\
	\mbox{subject\,to\,}\phantom{=} & \bm{a}\left(\frac{\Delta}{2}\right)^{\ast}\diag(\bar{\bm{g}})\bm{p}= e^{i\theta/2}, \nonumber \\
	& \bm{a}\left(-\frac{\Delta}{2}\right)^{\ast}\diag(\bar{\bm{g}})\bm{p}=e^{-i\theta/2},\nonumber \\
	& \dot{\bm{a}}\left(\frac{\Delta}{2}\right)^{\ast}\diag(\bar{\bm{g}})\bm{p}=0,\nonumber \\
	& \dot{\bm{a}}\left(-\frac{\Delta}{2}\right)^{\ast}\diag(\bar{\bm{g}})\bm{p}=0.
\end{align}
A key property of $\bm{p}_V$, recalled in the following lemma, is its equivalence with the minimal norm solution $\hat{\bm{p}}_{\min}$ under additional assumptions.
\begin{lem}[Equivalence of the minimal vanishing derivative solution \cite{duval_exact_2015}]\label{lem:equivalence}
If the solution $\bm{p}_V$ of \eqref{eq:minimalVanishingPreCertificate}  satisfies $\left\Vert \Phi_{g}^{\ast}\left(\bm{p}_V\right) \right\Vert_{\infty} \leq 1$, then $\bm{p}_V = \hat{\bm{p}}_{\min}$ is equal to the minimum norm solution for the measure $\mu_{\star}$.
\end{lem}
Let $\bm{q}_V  = \diag\left(\bar{\bm{g}}\right)\bm{p}_V\in\mathbb{C}^N$, and the associated polynomial $Q_V(t) = \bm{a}(t)^* \bm{q}_V$ be the minimal vanishing derivative polynomial.
Note that by construction, $\bm{Q}_V(t)$ verifies the interpolation constraints \eqref{eq:dualCertificateConditions-interp1} and \eqref{eq:dualCertificateConditions-interp2}. Additionally, if $\bm{q}_V$ also verifies the extremal constraint \eqref{eq:dualCertificateConditions-extremalConstraint}, $\left\Vert \Phi_{g}^{\ast}\left(\bm{p}_V\right) \right\Vert_{\infty}   = \sup_{t\in\mathbb{T}} \left\vert Q_V(t) \right\vert \leq 1$, then $\bm{p}_V = \hat{\bm{p}}_{\min}$ by Lemma \ref{lem:equivalence}. Therefore, it is enough to show that $Q_V(t)$ verifies \eqref{eq:dualCertificateConditions-extremalConstraint} and \eqref{eq:nonDegenerateSourceCondition} to conclude on the support stability of $\mu_\star$ for the sampling operator $\Phi_{g}$. To ensure that those conditions can be met, we start by studying the asymptotic of $Q_V(t)$ when $N \to \infty$.

\subsection{Asymptotic analysis of \texorpdfstring{$Q_V(t)$}{QV(t)}}
We start  the asymptotic analysis of the trigonometric polynomial $Q_V(t) = \bm{a}(t)^{\ast}\bm{q}_V$ by defining the discrete auto-correlation function $K(t)$ of the PSF $g(\tau)$ as
\begin{equation}
\label{eq:definitionOfK}
	K(t) \triangleq \sum_{k=-n}^{n} \left\vert g_{k}  \right\vert^2 e^{i2\pi k t} = \sum_{k=-n}^{n} \left\vert G\left(\frac{Bk}{N} \right) \right\vert^2 e^{i2\pi k t}, \quad \forall t\in\mathbb{T},
\end{equation}
which is a real and even trigonometric polynomial. The following lemma, whose proof is delayed to Appendix~\ref{sec:ProofOfExpressionInterpolationPolynomial}, gives an explicit expression of the polynomial $Q(t)$ in terms of the parameters $\Delta$, $\theta$ and the polynomial $K(t)$.
\begin{lem}
\label{lem:expressionInterpolationPolynomial}
Suppose that $\bm{g}$ has at least four non-zero coefficients, then for all $\Delta \in (0,\frac{1}{2}]$ and all $\theta \in [0,\pi]$, $Q_V$ can be decomposed as
\begin{equation}
	\label{eq:expressionOfQV}
	Q_V(t) = \cos\left(\frac{\theta}{2}\right) R_{\Delta}(t) + i \sin\left(\frac{\theta}{2}\right) S_{\Delta}(t),\quad \forall t \in\mathbb{T},
\end{equation}
where $R_{\Delta}$ and $S_{\Delta}$ are respectively the even and odd real trigonometric polynomials, independent of $\theta$, given by
\begin{subequations}
\begin{align}
		R_{\Delta}(t) &= C_{R}(\Delta)^{-1}\left[ \left(-K^{\pprime}(0) + K^{\pprime}(\Delta)\right)\left(K(t-\frac{\Delta}{2}) + K(t+\frac{\Delta}{2}) \right)
		+  K^{\prime}(\Delta)\left(K^{\prime}(t-\frac{\Delta}{2}) - K^{\prime}(t+\frac{\Delta}{2})\right)\right], \\
		S_{\Delta}(t) &= C_{S}(\Delta)^{-1}\left[ \left(-K^{\pprime}(0) -K^{\pprime}(\Delta)\right)\left(K(t-\frac{\Delta}{2}) - K(t+\frac{\Delta}{2}) \right)
		-  K^{\prime}(\Delta)\left(K^{\prime}(t-\frac{\Delta}{2}) + K^{\prime}(t+\frac{\Delta}{2})\right)\right],
		\end{align}
\end{subequations}
where the quantities $C_{R}(\Delta),C_{S}(\Delta)$ are positive and for all $\Delta \in (0,\frac{1}{2}]$ given as
\begin{subequations}
	\begin{align}
		C_{R}(\Delta) &= \left(-K^{\pprime}(0) + K^{\pprime}(\Delta)\right)\left(K(0)+K(\Delta)\right) - {K^\prime}(\Delta)^2 > 0,\\
		C_{S}(\Delta) &= \left(-K^{\pprime}(0) - K^{\pprime}(\Delta)\right)\left(K(0)-K(\Delta)\right) - {K^\prime}(\Delta)^2 > 0.
	\end{align}
\end{subequations}
\end{lem}

Next, we demonstrate in Lemma \ref{lem:convergenceOfK} the uniform convergence of $K$ towards the auto-correlation function~$\kappa$.
\begin{lem}[Uniform convergence of $K$]
\label{lem:convergenceOfK} Under the hypothesis of Theorem \ref{thm:stableResolutionLimit},
\begin{equation}
\lim_{N\to \infty} \sup_{t\in\mathbb{T}} \left\vert \left(\frac{B}{N}\right)^{\ell + 1} K^{(\ell)} \left(t\right) - \kappa^{(\ell)}\left(\frac{Nt}{B}\right) \right\vert = 0, \quad \ell = 0,1,2,3.
\end{equation}
\end{lem}
The proof of the above is presented in Appendix \ref{sec:ProofConvergenceOfK}.  By (H1) and (H2), the Fourier transform $G$ of the PSF $g$ is non-zero on a non-empty open interval $I\subset \mathbb{R}$. By (H2), $G$ is band-limited within $B$, and one must have $I\subset (-B/2,B/2)$. It comes that if $N \geq \left\lceil\frac{4}{\vert I \vert} \right\rceil$, at least four elements of the form $\left\{kB/N\right\}_{{\vert k \vert}\leq n}$ will fall into $I$, and the vector $\bm{g}$ has at least four non-zero coefficients. Lemma \ref{lem:convergenceOfK} can be applied to the expression of $Q_V$ \eqref{eq:expressionOfQV} to get the existence of a function $\mathcal{Q}_{V}$ verifying the convergence
\begin{equation}
\label{eq:uniformConvergeceOfQ}
\lim_{N\to\infty} \sup_{t\in\mathbb{T}} \left\vert \left(\frac{B}{N}\right)^{\ell} Q_V^{(\ell)}(t) - \mathcal{Q}_{V}^{(\ell)}\left(\frac{Nt}{B}\right)
\right\vert = 0, \quad \ell=0,1,2.
\end{equation}
Moreover, defining $\beta = N \Delta / B$, the limit function $\mathcal{Q}_{V}$ writes as
\begin{equation}
	\label{eq:qVExpression}
	\mathcal{Q}_{V}(\tau) = \cos\left(\frac{\theta}{2}\right)r_{\beta}(\tau) + i \sin\left(\frac{\theta}{2}\right)s_{\beta}(\tau), \quad \forall\tau\in\mathbb{R}.
\end{equation}
The intermediate real functions $r_\beta,s_\beta$ in the above expression are even and respectively, given by the limits of $R_{B\beta/N}(B\tau/N)$ and $S_{B\beta/N}(B\tau/N)$, so that
\begin{subequations}
\label{eq:rAndSExpression}
	\begin{align}
		r_\beta(\tau) &= C_{r}(\beta)^{-1}\left[ \left(-\kappa^{\pprime}(0) + \kappa^{\pprime}(\beta)\right) u_{\beta}(\tau) +
		\kappa^{\prime}(\beta)v_{\beta}^\prime(\tau) \right], \\
		s_\beta(\tau) &= C_{s}(\beta)^{-1}\left[ \left(-\kappa^{\pprime}(0) -\kappa^{\pprime}(\beta)\right)v_{\beta}(\tau)
		- \kappa^{\prime}(\beta)u_{\beta}^\prime(\tau)\right],
	\end{align}
\end{subequations}
where the function $u_\beta,v_\beta$ are defined in \eqref{eq:uAndvDefinition}, and  $C_{r}(\beta),C_{s}(\beta)$ are positive quantities given by
\begin{subequations}
\label{eq:deltaRAndDeltaSExpression}
	\begin{align}
		C_{r}(\beta) &= \left(-\kappa^{\pprime}(0) + \kappa^{\pprime}(\beta)\right)\left(\kappa(0)+\kappa(\beta)\right) - {\kappa^\prime}(\beta)^2 > 0,\\
		C_{s}(\beta) &= \left(-\kappa^{\pprime}(0) - \kappa^{\pprime}(\beta)\right)\left(\kappa(0)-\kappa(\beta)\right) - {\kappa^\prime}(\beta)^2 > 0.
	\end{align}
\end{subequations}

We highlight that the compositions \eqref{eq:expressionOfQV} and \eqref{eq:qVExpression} indicate that both $Q_V(t)$  and $\mathcal{Q}_V(\tau)$ are Hermitian:  $\overline{Q_V\left(t\right)} = Q_V\left(-t\right)$ for all $t\in\mathbb{T}$ and $\overline{\mathcal{Q}_V\left(\tau\right)} = \mathcal{Q}_V\left(-\tau\right)$ for all $\tau\in\mathbb{R}$. Suppose that the limit function $\mathcal{Q}_V$ verifies the conditions
\begin{subequations}
\label{eq:conditionOnLimitQ}
    \begin{align}
    \label{eq:conditonOnLimitQ-concavity}
    \frac{{\rm d}^{2}\left|{\mathcal{Q}_{V}}\right|}{{\rm d}\tau^{2}}\left(\frac{\beta}{2}\right) &< 0 ,\\
    \label{eq:conditonOnLimitQ-modulus}
    \left\vert \mathcal{Q}_{V}(\tau) \right\vert &< 1, \quad \forall \tau \in \mathbb{R}^{+}\backslash\left\{\frac{\beta}{2}\right\}.
    \end{align}
\end{subequations}
It follows from the uniform convergence \eqref{eq:uniformConvergeceOfQ} of $Q_V$ towards $\mathcal{Q}\left(Nt/B\right)$, and the Hermitian symmetry of $Q_V$ and $\mathcal{Q}_V$ that there must exist an integer $\widetilde{N}_0$ so that for all $N \geq \widetilde{N}_0$, $Q_V$ verifies \eqref{eq:dualCertificateConditions-extremalConstraint} and \eqref{eq:nonDegenerateSourceCondition}.

Therefore, using the results of Section \ref{subsec:Minimal-vanishing-pre-certificate}, if $\mathcal{Q}_V$  verifies \eqref{eq:conditionOnLimitQ}, then $\mu_\star\left(\Delta,\theta\right)$ verifies the non-degenerate source condition with respect to the measurement operator $\Phi_g$ whenever $N \geq N_0 \triangleq \max\left\{\left\lceil\frac{4}{\vert I \vert} \right\rceil, \widetilde{N}_0 \right\}$.

\subsection{Non-degeneracy of \texorpdfstring{$\mu_\star(\Delta,\theta)$}{μ(ϵ,θ)}} \label{subsec:nondegeneracyOfMu}

It remains to derive sufficient conditions under which the limit  function $\mathcal{Q}_{V}$ satisfies \eqref{eq:conditionOnLimitQ} to conclude on the desired result. First of all, it comes immediately from \eqref{eq:expressionOfQV} that
\begin{equation}
    \label{eq:QLimit-modulusSqaure}
	\vert \mathcal{Q}_{V}(\tau)\vert^2 = \cos^2\left(\frac{\theta}{2}\right)\vert r_\beta(\tau) \vert^2 + \sin^2\left(\frac{\theta}{2}\right)\vert s_\beta(\tau) \vert^2, \quad \forall \tau \in  \mathbb{R}.
\end{equation}

We start by discussing the condition \eqref{eq:conditonOnLimitQ-modulus}. From \eqref{eq:QLimit-modulusSqaure}, it is clear that  \eqref{eq:conditonOnLimitQ-modulus} is verified for every  $\theta$ in $[0,\pi]$ if and only if $\vert r_\beta(\tau)  \vert < 1$ and $\vert s_\beta(\tau)  \vert < 1$  for all $\tau$ in $\mathbb{R}^{+}\backslash\left\{\beta/2\right\}$. Those last inequalities can be equivalently interpreted in terms of the functions $\tilde{r}_\beta(\tau)$, $\tilde{s}_\beta(\tau)$ introduced in \eqref{eq:rtildeAndStilde} from the identity  $\tilde{r}_\beta(\tau) = C_{r}(\beta)r_\beta(\tau)$ and $\tilde{s}_\beta(\tau) = C_{s}(\tau)s_\beta(\tau)$ for all $\tau\in\mathbb{R}$, as we have that
\begin{subequations}
\begin{align*}
\left\vert r_\beta(\tau) \right\vert < 1 &\Longleftrightarrow \frac{\left\vert \tilde{r}_\beta(\tau) \right\vert}{C_{r}(\beta)} < 1,\\
\left\vert s_\beta(\tau) \right\vert < 1 &\Longleftrightarrow \frac{\left\vert \tilde{s}_\beta(\tau) \right\vert}{C_{s}(\beta)} < 1.
\end{align*}
\end{subequations}
Hence, the condition \eqref{eq:conditonOnLimitQ-modulus} is equivalent to
\begin{subequations}
\label{eq:rAndSPositiveConditions}
    \begin{align}
	C_{r}(\beta) - \left\vert \tilde{r}_\beta(\tau) \right\vert &> 0,\quad \forall \tau \in \mathbb{R}^{+} \backslash \left\{ \frac{\beta}{2}\right\} \label{eq:RpositiveCond}\\
	C_{s}(\beta) - \left\vert \tilde{s}_\beta(\tau) \right\vert &> 0,\quad \forall \tau \in \mathbb{R}^{+} \backslash \left\{ \frac{\beta}{2}\right\} \label{eq:SpositiveCond}.
\end{align}
\end{subequations}
From the definitions  \eqref{eq:rtildeAndStilde} and \eqref{eq:deltaRAndDeltaSExpression}, one has  $C_{r}(\beta) =  \tilde{r}_\beta(\frac{\beta}{2})$ and $C_{s}(\beta) =  \tilde{s}_\beta(\frac{\beta}{2})$. Leveraging the hypothesis $\beta = N \Delta/B  > \max \left\{ \gamma^{\star}_1, \gamma^{\star}_2\right\} / B$, the functions $C_{r}(\beta) - \left\vert \tilde{r}_\beta(\tau) \right\vert$ and $C_{s}(\beta) - \left\vert \tilde{s}_\beta(\tau) \right\vert$ are non-negative and can reach $0$ only for $\tau = \beta/2$. Hence \eqref{eq:rAndSPositiveConditions} and subsequently \eqref{eq:conditonOnLimitQ-modulus}
 are both verified under the hypothesis of Theorem \ref{thm:stableResolutionLimit}.

It remains to show that $\mathcal{Q}_V$ satisfies \eqref{eq:conditonOnLimitQ-concavity}. We recall that the second derivative of the modulus of a function $q = q_R + iq_I$, where $q_R,q_I$ are its real and imaginary part, respectively, reads (see e.g. \cite{tang_compressed_2013})
\begin{equation}
	\label{eq:modulusDerivativeExpression}
	\frac{{\rm d}^{2}\left|{q}\right|}{{\rm d}\tau^{2}}\left(\tau\right)=-\frac{\left(q_R\left(\tau\right)q_R^{\prime}\left(\tau\right)+q_I\left(\tau\right)q_I^{\prime}\left(\tau\right)\right)^{2}}{\left|q\left(\tau\right)\right|^{3}}+\frac{\left|q^{\prime}\left(\tau\right)\right|^{2}+q_R\left(\tau\right)q_R^{\pprime}\left(\tau\right)+q_I\left(\tau\right)q_I^{\pprime}\left(\tau\right)}{\left|q\left(\tau\right)\right|}.
\end{equation}
Evaluating the above at $\tau=\frac{\beta}{2}$ for the function $\mathcal{Q}_V$ leads to
\begin{align}\label{eq:secondDerivative}
	\frac{{\rm d}^{2}\left|\mathcal{Q}_V\right|}{{\rm d}\tau^{2}}\left(\frac{\beta}{2}\right) & =\cos\left(\frac{\theta}{2}\right) r^{\pprime}_{\beta}\left(\frac{\beta}{2}\right)+\sin\left(\frac{\theta}{2}\right)s^{\pprime}_{\beta}\left(\frac{\beta}{2}\right) \nonumber \\
	&= \left(-{\kappa^{\pprime}(0)}^2 + {{\kappa^{\pprime}}(\beta)}^2 - \kappa^{\prime}(\beta)\kappa^{\ppprime}(\beta)\right)\left( C_{r}(\beta)^{-1} \cos\left(\frac{\theta}{2}\right) + C_{s}(\beta)^{-1}\sin\left(\frac{\theta}{2}\right)\right),
\end{align}
which is strictly negative for every $\theta\in \left[0,\pi\right]$ by the hypothesis $\beta = N \Delta/B > \gamma_3^{\star}/B$.

As a result, under the hypothesis of Theorem \ref{thm:stableResolutionLimit}, if $\Delta = B \beta / N \geq \gamma^{\star}/N$ with $\gamma^{\star} = \max \left\{ \gamma^{\star}_1, \gamma^{\star}_2, \gamma^{\star}_3 \right\}$, there exists $N_0\in\mathbb{N}$ such that for every $N \geq N_0$ the canonical measure $\mu_{\star}(\Delta,\theta)$ will verify the non-degenerate source condition as in Def. \ref{def:nonDegenerateSourceCondition} for the sampling operator $\Phi_{g}$ for every $\theta \in [0,\pi]$. We conclude on the desired statement by an application of Theorem \ref{prop:supportStability}. \qed

\section{Extension to Multiple Point Sources} \label{sec:extensions}

The stable resolution limit presented in Section~\ref{sec:results} only applies to the simple case when only two point sources are present. In this section, we show that the same resolution limit continues to govern the support stability in a multi-source setting including two close-located sources and other well-separated sources.

We consider an extension of the normalized measure $\mu_\star$ in \eqref{eq:normalizedMeasure} composing $S$ point sources at locations $\mathcal{T} = \left\{t_s\right\}_{s=1}^{S} \subset \mathbb{T}$, given as
\begin{equation}\label{eq:normalizedMeasureSSources}
	\mu_\star(t) = \sum_{s=1}^S c_s \delta(t-t_s),
\end{equation}
where $\left\{c_s \in\mathbb{C}\right\}_{s=1}^{S}$ are non-zero complex amplitudes. Similarly, we can extend the notion of support stability, defined in Def.~\ref{def:support_stability} to include more spikes. Without loss of generality, we assume that $t_1$ and $t_2$ are closely located, while the other sources are well-separated from each other. Theorem \ref{thm:multipleSources} establishes that the support stability of the Beurling-LASSO estimator can be ensured under the same separation condition between the two close sources as that of Theorem \ref{thm:stableResolutionLimit}, depending only on the characteristics of the PSF.

\begin{thm}[Extension to multiple point sources]\label{thm:multipleSources} Suppose that the PSF $g$ satisfies the  regularity conditions (H1)-(H4) of Theorem \ref{thm:stableResolutionLimit}. Let $S \geq 3$ be a constant that does not grow with $N$, and the support set $\mathcal{T} = \left\{t_s\right\}_{s=1}^{S}$ of $\mu_{\star}$ verifies
		\begin{subequations} \label{eq:multiSources-separationCondition}
			\begin{align}
				\Delta:=\left\vert t_2 - t_1 \right\vert_\mathbb{T} & > \gamma^\star  / N, \\
				\left\vert t_{s^{\prime}} - t_{s} \right\vert_\mathbb{T} &= \omega(1/N),  \quad \forall s,s^\prime \mbox{ with } s \neq s^\prime \mbox{ and } \{s,s^\prime\}\neq \{1,2\},
			\end{align}
		\end{subequations}
	where $\gamma^\star$ is the constant defined in the statement of Theorem \ref{thm:stableResolutionLimit}. Then there exists  $N_0\in\mathbb{N}$ such that, for every $N\geq N_0$ and every measure $\mu_{\star}$ of the form \eqref{eq:normalizedMeasureSSources} satisfying \eqref{eq:multiSources-separationCondition}, there exists $\alpha > 0$ such that the Beurling-LASSO estimator $\hat{\mu}_\lambda\left(\Phi_{g}\left(\mu_\star\right) + \bm{w} \right)$ with the regularization parameter $\lambda = \alpha^{-1}\left\Vert \bm{w} \right\Vert_2$  is support stable.
\end{thm}

The proof of this theorem is given in Appendix \ref{sec:proofOfMultipleSources}. Similar to the proof of Theorem \ref{thm:stableResolutionLimit} outlined in Section~\ref{sec:proofs}, it relies on the characterization of the non-degenerate source condition presented in \cite{duval_exact_2015}. More precisely, we show that, under the assumptions of Theorem \ref{thm:multipleSources}, the minimal vanishing derivative polynomial associated to the measure $\mu_\star$ with $S$ sources as in \eqref{eq:normalizedMeasureSSources} has the same asymptotic behavior as the sum of the polynomial associated with the measure with two close spikes, whose properties were studied in Section \ref{sec:proofs}, and the polynomials associated with each well-separate single spike.

\begin{figure}[ht]
	\centering
	\begin{tabular}{ccc}
	\includegraphics[width=0.31\textwidth]{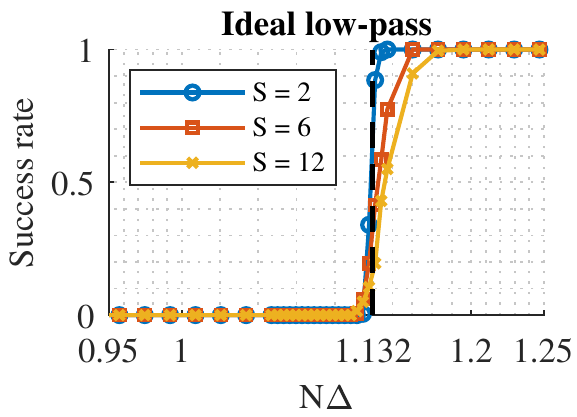}
	&
	\includegraphics[width=0.31\textwidth]{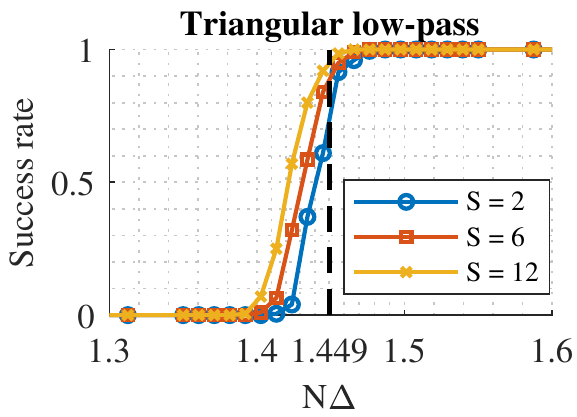}
	&
	\includegraphics[width=0.31\textwidth]{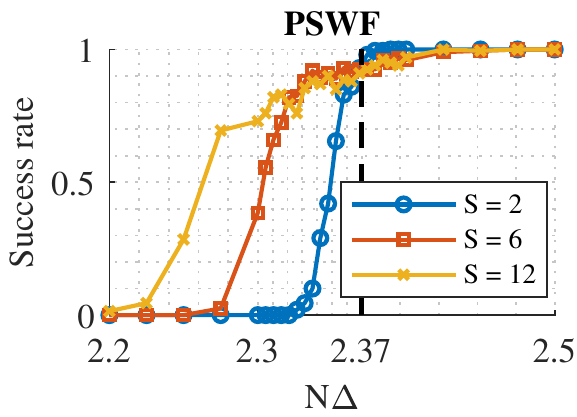}
	\\
	(a) $g(\tau) = \sinc(\pi \tau)$ & (b) $g(\tau) = {\sinc(\pi \tau/2)}^2$ & (c) $g(\tau) = \psi_{5}(\tau)$
	\end{tabular}
	\caption{Empirical success rate for the Beurling-LASSO estimator to return a measure with $S$ point sources, for three different point spread functions, under additive white Gaussian noise, as a function of the separation parameter $N\Delta$. The support stability threshold $\gamma^{\star}$ predicted by Theorem \ref{thm:multipleSources} is shown in a black dashed line. Here, we set $N=101$, $\mathrm{SNR}=60$dB and $\left\vert t_{s^{\prime}} - t_{s} \right\vert_\mathbb{T} = 5/N$ as a separation for the well separated spikes $s \neq s^\prime$ and $\{s,s^\prime\}\neq \{1,2\}$. The results are averaged over 200 trials.}
	\label{fig:successRate_multiSpikes}
\end{figure}

We compare in Fig.~\ref{fig:successRate_multiSpikes} the theoretical threshold $\gamma^\star$ provided by Theorem \ref{thm:multipleSources} with the empirical success rate of the Beurling-LASSO estimator to output a measure $\hat{\mu}$ with $S$ spikes given a ground truth $\mu_{\star}$ of the form \eqref{eq:normalizedMeasureSSources}, for different values of the separation parameter $\Delta$ and for different PSF $g(\tau)$. In the experiments, we set $\left\vert t_{s^{\prime}} - t_{s} \right\vert_\mathbb{T} = 5/N$ as a separation for the well-separated spikes, where $s \neq s^\prime$ and $\{s,s^\prime\}\neq \{1,2\}$. The amplitude of the colliding spikes $c_1,c_2$ are chosen with opposite signs $c_1 = 1$, $c_2=-1$, while the amplitudes of the well-separated ones $\{ c_s \}_{s=3}^{S}$ are drawn uniformly at random over the complex unit circle.

\section{Conclusions} \label{sec:conclusions}

This paper studies the support stability of the Beurling-LASSO estimator for estimating two closely located point sources, possibly in the presence of a finite number of well-separated point sources, and characterizes the resolution limit as a function of the PSF, above which the Beurling-LASSO estimator is support stable. Our result highlights and quantifies the role of PSF in noisy super resolution, which is not evident in the study of the noiseless setting. Our analysis is based on verifying the non-degenerate source conditions put forth in \cite{duval_exact_2015}. We believe it is possible to develop similar stable resolution limits in higher dimensions \cite{candes_towards_2014,chi_compressive_2015}, by carefully generalizing the arguments in our paper, which is left for the future work. Finally, it will be interesting to study how the availability of multiple snapshots impact the stable resolution limit \cite{li2015off}.

\appendix

\section{Proof of Lemma \ref{lem:invarianceOfNonDegenerateSourceCondition}}\label{sec:ProofOfInvarianceOfNonDegenerateSourceCondition}

We restrict the proof to showing the translation invariance of the non-degenerate source condition. The scaling and time-reversal invariances can be demonstrated by following an analogous reasoning.

Consider a measure $\mu_1(t) = c_1 \delta (t - t_{1,1}) + c_2\delta (t-t_{1,2})$ for all $t\in\mathbb{T}$, and let
$$\mu_2(t) = \mu_1(t-t_0) = c_1 \delta (t - t_{2,1}) + c_2\delta (t-t_{2,2}),$$
for some $t_0 \in \mathbb{T}$ with $t_{2,1} = t_{1,1} -t_0$ and $t_{2,2} = t_{1,2} -t_0$. Moreover, let $\bm{x}_j = \Phi_g(\mu_j)$ for $j=1,2$, and consider an element $\bm{p}_1$ within the feasible set of \eqref{eq:noisyDual}. For any $t_0\in\mathbb{T}$, we let $\bm{p}_2 = \bm{a}(-t_0) \odot \bm{p}_1$. Since $\bm{p}_1\in \mathcal{D}_0(\bm{x}_1)$ by hypothesis, we have that
\begin{align}
\label{eq:equivalence-feasability}
\left\Vert {\Phi_g}^{\ast} \left( \bm{p}_2 \right) \right\Vert_{\infty} &= \sup_{t\in\mathbb{T}} \left\vert \bm{a}(t)^* \diag(\bar{\bm{g}}) \bm{p}_2 \right\vert  \nonumber\\
&= \sup_{t\in\mathbb{T}} \left\vert \bm{a}(t)^{\ast} \diag(\bar{\bm{g}}) \left( \bm{a}(-t_0) \odot \bm{p}_1 \right) \right\vert \nonumber \\
&= \sup_{t\in\mathbb{T}} \left\vert \bm{a}(t+t_0)^{\ast} \diag(\bar{\bm{g}})  \bm{p}_1 \right\vert \nonumber\\
&= \sup_{t\in\mathbb{T}} \left\vert \bm{a}(t)^{\ast} \diag(\bar{\bm{g}})  \bm{p}_1 \right\vert =
\left\Vert \Phi_g^{\ast} \left( \bm{p}_1 \right)  \right\Vert_{\infty} \leq 1.
\end{align}
It comes that $\bm{p}_2$ also lies in the feasible set of \eqref{eq:noisyDual}. Moreover, from the definition \eqref{eq:SamplingOperator} of the sampling operator $\Phi_{g}$, we have that
$$
\bm{x}_2 = \Phi_{g}(\mu_2) = \Phi_{g}\left(\mu_1(\cdot - t_0)\right) = \bm{a}(-t_0) \odot \bm{x}_1.
$$
Evaluating the cost function of \eqref{eq:noisyDual} for $\bm{p}_2$ yields
\begin{align*}
    \left\langle \bm{x}_2, \bm{p}_2 \right\rangle_{\mathbb{R}} = \left\langle \bm{a}(-t_0) \odot \bm{x}_1, \bm{a}(-t_0) \odot \bm{p}_1 \right\rangle_{\mathbb{R}} = \left\langle \bm{x}_1, \bm{p}_1 \right\rangle_{\mathbb{R}}.
\end{align*}
This implies that $\bm{p}_2 \in \mathcal{D}_0(\bm{x}_2)$ is a solution of \eqref{eq:noisyDual} with an input $\bm{x}_2$ if and only if $\bm{p}_2 \in \mathcal{D}_0(\bm{x}_1)$ is a solution of \eqref{eq:noisyDual} with an input $\bm{x}_1$. Further noticing that $\left\Vert \bm{p}_2 \right\Vert_2 =  \left\Vert \bm{p}_1 \right\Vert_2$, we conclude that the minimal norm elements $\hat{\bm{p}}_{1,\min}$, $\hat{\bm{p}}_{2,\min}$ of the sets $\mathcal{D}_0(\bm{x}_1)$, $\mathcal{D}_0(\bm{x}_2)$, respectively, are linked by the relation
\[
\hat{\bm{p}}_{2,\min} = \bm{a}(-t_0) \odot \hat{\bm{p}}_{1,\min}.
\]

Now, suppose that $\mu_1$ verifies the non-degenerate condition with respect to $\Phi_g$, and define the trigonometric polynomials $\hat{Q}_{j,\min}(t) = \bm{a}(t)^{\ast}\diag(\bar{\bm{g}})\hat{\bm{p}}_{j,\min}$, for $j=1,2$. By Def. \ref{def:nonDegenerateSourceCondition}, we have that
\begin{align}
\hat{Q}_{2,\min}(t - t_0) &= \bm{a}(t-t_0)^{\ast}\diag(\bar{\bm{g}})\hat{\bm{p}}_{2,\min} \nonumber \\
&= \bm{a}(t-t_0)^{\ast}\diag(\bar{\bm{g}})\left( \bm{a}(-t_0) \odot \hat{\bm{p}}_{1,\min}\right)\nonumber \\
&= \bm{a}(t)^{\ast}\diag(\bar{\bm{g}})\hat{\bm{p}}_{1,\min} = \hat{Q}_{1,\min}(t), \qquad \forall t\in\mathbb{T}.
\end{align}
Therefore $\hat{Q}_{2,\min}(t_{2,s}) = \hat{Q}_{2,\min}(t_{1,s}) = \sign(c_s)$ for $s=1,2$, and   $\vert \hat{Q}_{2,\min}(t)\vert < 1$ for all $t\in\mathbb{T}\backslash\left\{t_{2,1},t_{2,2}\right\}$. Consequently $\hat{Q}_{2,\min}(t)$  verifies conditions \eqref{eq:dualCertificateConditions}. A similar reasoning also yields
$$
\frac{{\rm d}^{2}}{{\rm d}t^{2}}\left|\hat{Q}_{2,\min} \left(t_{2,s}\right) \right| = \frac{{\rm d}^{2}}{{\rm d}t^{2}}\left|\hat{Q}_{1,\min} \left(t_{1,s}\right) \right|  <0, \quad s = 1,2,
$$
and $\hat{Q}_{2,\min}(t)$ verifies~\eqref{eq:nonDegenerateSourceCondition}. We conclude that $\mu_2$ verifies the non-degenerate source condition of the sampling operator $\Phi_g$, completing the proof of the translation invariance. \qed

\section{Proof of Lemma \ref{lem:expressionInterpolationPolynomial}}\label{sec:ProofOfExpressionInterpolationPolynomial}

We derive the expression of the polynomial $Q_V(t) = \bm{a}(t)^*\diag(\bar{\bm{g}}) \bm{p}_V$. By the orthogonal projection theorem, the solution $\bm{p}_V$  of the quadratic program \eqref{eq:minimalVanishingPreCertificate} belongs to the subspace
\[
	\bm{p}_V \in \spann\left\{ \diag(\bm{g})\bm{a}\left(\frac{\Delta}{2}\right), \diag(\bm{g})\bm{a}\left(-\frac{\Delta}{2}\right), \diag(\bm{g})\dot{\bm{a}}\left(\frac{\Delta}{2}\right), \diag(\bm{g})\dot{\bm{a}}\left(-\frac{\Delta}{2}\right) \right\}.
\]
Recalling that $\bm{q}_V = \diag(\bar{\bm{g}})\bm{p}_V$, it yields
\[
	\bm{q}_V \in \spann\left\{ \diag(\vert\bm{g}\vert^2)\bm{a}\left(\frac{\Delta}{2}\right), \diag(\vert\bm{g}\vert^2)\bm{a}\left(-\frac{\Delta}{2}\right), \diag(\vert\bm{g}\vert^2)\dot{\bm{a}}\left(\frac{\Delta}{2}\right), \diag(\vert\bm{g}\vert^2)\dot{\bm{a}}\left(-\frac{\Delta}{2}\right) \right\},
\]
where, with a slight abuse of notation, the operator $\vert \cdot \vert^2$ is interpreted element-wise. Remarkably, denoting by $K(t) = \sum_{k=-n}^{n} \left\vert g_k \right\vert ^2 e^{i 2\pi k t}$, we have that, for every $t,t^{\prime} \in \mathbb{T}$.
\begin{align*}
    \bm{a}\left(t\right)^{\ast}\diag(\vert\bm{g}\vert^2)\bm{a}\left(t^\prime\right) &= \sum_{k=-n}^n \vert g_k \vert^2 e^{2\pi i k (t - t^\prime)} = K(t - t^{\prime})\\
    \bm{a}\left(t\right)^{\ast}\diag(\vert\bm{g}\vert^2)\dot{\bm{a}}\left(t^\prime\right) &= - 2\pi i \sum_{k=-n}^{n} k \left\vert g_k \right\vert^2 e^{2\pi i k (t - t^{\prime})} = - K^{\prime}(t-t^{\prime}).
\end{align*}
Therefore the polynomial $Q_V(t) = \bm{a}(t)^{\ast} \bm{q}_V$ lies in the span of the translations of the auto-correlation $K(t)$ and its derivative $K^{\prime}(t)$ at the spike locations $\{-\Delta/2,\Delta/2\}$. Introducing a change of basis for the purpose of convenience, the polynomial $Q_V(t)$ writes for all $t\in\mathbb{T}$ as
\begin{multline}
	Q_V(t) = \alpha_R \left( K(t-\frac{\Delta}{2}) + K(t+\frac{\Delta}{2}) \right) - \beta_R \left( K^{\prime}(t-\frac{\Delta}{2}) - K^{\prime}(t+\frac{\Delta}{2}) \right) \\
	+ i\alpha_S \left( K(t-\frac{\Delta}{2}) - K(t+\frac{\Delta}{2}) \right) - i\beta_S \left( K^{\prime}(t-\frac{\Delta}{2}) + K^{\prime}(t+\frac{\Delta}{2}) \right),
\end{multline}
for some coefficients $\left\{\alpha_R,\beta_R,\alpha_S,\beta_S\right\}\subset\mathbb{C}$, which are fully determined by the four interpolation constraints of \eqref{eq:minimalVanishingPreCertificate}, that can be reinterpreted as
\begin{subequations}
    \label{eq:QV-linearSystem}
  \begin{align}
	\frac{1}{2}\left(Q_V(\frac{\Delta}{2}) + Q_V(-\frac{\Delta}{2})\right) &= \cos\left(\frac{\theta}{2}\right),\\
	\frac{1}{2}\left(Q^{\prime}_V(\frac{\Delta}{2}) - Q^{\prime}_V(-\frac{\Delta}{2})\right) &= 0,\\
	\frac{1}{2i}\left(Q_V(\frac{\Delta}{2}) - Q_V(-\frac{\Delta}{2})\right) &= \sin\left(\frac{\theta}{2}\right),\\
	\frac{1}{2i}\left(Q^{\prime}_V(\frac{\Delta}{2}) + Q^{\prime}_V(-\frac{\Delta}{2})\right) &= 0.
\end{align}
\end{subequations}
The linear system \eqref{eq:QV-linearSystem} can be reformulated in terms of the $4 \times 4$ block diagonal system
\begin{equation}\label{eq:linearInterpolationSystem}
	\begin{bmatrix}
		\bm{M}_{R} & \bm{0} \\
		\bm{0} 		 & \bm{M}_S
	\end{bmatrix}
	\begin{bmatrix}
		\alpha_R \\ \beta_R \\ \alpha_S \\ \beta_S
	\end{bmatrix}
	=
	\begin{bmatrix}
		\cos\left(\frac{\theta}{2}\right) \\ 0 \\ \sin\left(\frac{\theta}{2}\right) \\ 0
	\end{bmatrix},
\end{equation}
where each of the diagonal blocks can be decomposed as
\begin{subequations}
\label{eq:M-Decomp}
   \begin{align}
	\bm{M}_R =
	\begin{bmatrix}
		K(0) + K(\Delta) & K^{\prime}(\Delta)\\
		K^{\prime}(\Delta) & -K^{\pprime}(0) + K^{\pprime}(\Delta)
	\end{bmatrix}
	&=
	\bm{A}_R^\ast \diag(\bm{g})^\ast \diag(\bm{g}) \bm{A}_R ,\\
	\bm{M}_S =
	\begin{bmatrix}
		K(0) - K(\Delta) & -K^{\prime}(\Delta)\\
		-K^{\prime}(\Delta) & -K^{\pprime}(0) - K^{\pprime}(\Delta)
	\end{bmatrix}
	&=
	\bm{A}_S^\ast \diag(\bm{g})^\ast \diag(\bm{g}) \bm{A}_S.
\end{align}
\end{subequations}

Here, the intermediate matrices $\bm{A}_R,\bm{A}_S\in\mathbb{C}^{N\times 2}$ are given by
\begin{subequations}
    \begin{align*}
	\bm{A}_R &= 2^{-1/2}\begin{bmatrix}
		\bm{a}(\frac{\Delta}{2}) + \bm{a}(-\frac{\Delta}{2}) & \dot{\bm{a}}(\frac{\Delta}{2}) - \dot{\bm{a}}(-\frac{\Delta}{2})
	\end{bmatrix} ,\\ \bm{A}_S &= 2^{-1/2}\begin{bmatrix}
		\bm{a}(\frac{\Delta}{2}) - \bm{a}(-\frac{\Delta}{2}) & \dot{\bm{a}}(\frac{\Delta}{2}) + \dot{\bm{a}}(-\frac{\Delta}{2})
	\end{bmatrix}.
\end{align*}
\end{subequations}

Using the invertibility properties of Vandermonde matrices, it can be verified that the matrix $\left[\bm{A}_R,\bm{A}_S\right]$ and all its submatrices of size $4\times4$ are of the maximal rank whenever $0 < \Delta \leq \frac{1}{2}$. Consequently, if $\bm{g}$ has four or more non-zero entries, the matrices $\diag(\bm{g})\bm{A}_R$ and $\diag(\bm{g}) \bm{A}_S$ are also of the maximal rank whenever $0 < \Delta \leq \frac{1}{2}$. We conclude, using the decomposition on the right-hand-side of \eqref{eq:M-Decomp}, that the matrices $\bm{M}_R,\bm{M}_S$ are positive definite Hermitian matrices, and that the linear system \eqref{eq:linearInterpolationSystem} has a unique solution. Solving this system leads to the desired expression of $Q_V$, where we let $C_{R} = \mbox{det}(\bm{M}_R) > 0$ and $C_{S} = \mbox{det}(\bm{M}_S) > 0$. \qed

\section{Proof of Lemma \ref{lem:convergenceOfK}}\label{sec:ProofConvergenceOfK}
Let by $\kappa = \mathcal{K}(g) \in L_2$ the autocorrelation of $g$. We start by recalling that, by the  Wiener-Khinchin theorem, $\mathcal{F}(\kappa) = \left\vert G(\cdot) \right\vert^{2} \in L_{2}$. It comes by (H2) and (H3) that the function $f \mapsto (i 2 \pi f)^\ell \mathcal{F}(\kappa)(f)$ is also band-limited within $B$ and bounded for any $\ell \in \mathbb{N}$. Therefore it is absolutely integrable over $\mathbb{R}$ for any $\ell \in \mathbb{N}$. We conclude that $\kappa$ possesses derivatives of all orders and that
\[
	\mathcal{F}(\kappa^{(\ell)}):\; f \mapsto (i 2 \pi f)^\ell \mathcal{F}(\kappa)(f), \quad \forall\ell \in \mathbb{N}.
\]

Fix an odd integer $N=2n+1$. Using the definition \eqref{eq:definitionOfK} of $K$, we have for $\ell = 0,1,2,3$ that
\begin{align}
\left(\frac{B}{N}\right)^{\ell + 1} K^{(\ell)}(t) &= \frac{B}{N} \sum_{k=-n}^{n} \left(i2\pi \frac{Bk}{N}\right)^{\ell} \left\vert G\left(\frac{Bk}{N}\right) \right\vert^{2} e^{i 2 \pi k t} \nonumber\\
&= \frac{B}{N} \sum_{k\in\mathbb{Z}} \left(i2\pi \frac{Bk}{N}\right)^{\ell} \left\vert G\left(\frac{Bk}{N}\right) \right\vert^{2} e^{i 2 \pi k t} \nonumber \\
&= \sum_{k\in\mathbb{Z}} \mathcal{F}\left(\kappa^{(\ell)}\left(\frac{N}{B}\left(\,t + \cdot \,\right)\right)\right)\left(k\right), \label{eq:Poisson-frequency}
\end{align}
where we used the fact that $G(\frac{Bk}{N}) = 0$ for all $\vert k \vert \geq n+1$ by the hypothesis (H2) in the second equality, and identified the expression with the Fourier transform of the derivatives of $\kappa$ in the last equality. Hence, the series in the last equality converges absolutely and uniformly in $t$ as it has finitely many non-zero terms. Moreover the series
\begin{equation}
\label{eq:Poisson-time}
\sum_{k\in \mathbb{Z}} \kappa^{(\ell)}\left(\frac{N}{B}\left(t+k\right)\right) = \kappa^{(\ell)}\left(\frac{N}{B} t\right)  +  \sum_{\substack{k\in\mathbb{Z} \\ k \neq 0}} \kappa^{(\ell)} \left(\frac{N}{B}\left(t+k\right)\right) < \infty
\end{equation}
converges uniformly for $\ell = 0,1,2,3$ as $\kappa^{(\ell)}$ is bounded and by the hypothesis (H4). Applying the Poisson summation formula between the right-hand-side of \eqref{eq:Poisson-frequency} and the left-hand-side of \eqref{eq:Poisson-time} yields
\begin{align}
\label{eq:poisson-application}
    \left(\frac{B}{N}\right)^{\ell + 1} K^{(\ell)}(t) &= \sum_{k\in\mathbb{Z}} \mathcal{F}\left(\kappa^{(\ell)}\left(\frac{N}{B}\left(t+\cdot \, \right)\right)\right)\left(k\right) \nonumber \\
    &= \sum_{k\in \mathbb{Z}} \kappa^{(\ell)}\left(\frac{N}{B}\left(t+k\right)\right).
\end{align}
Substracting the term of index 0 in \eqref{eq:poisson-application}, taking the absolute value and the supremum over $t$ on both sides of the equality yields
$$
\sup_{t\in\mathbb{T}} \left\vert \left(\frac{B}{N}\right)^{\ell + 1} K^{(\ell)} \left(t\right) - \kappa^{(\ell)}\left(\frac{Nt}{B}\right) \right\vert = \sup_{t\in\mathbb{T}} \left\vert \sum_{\substack{k\in\mathbb{Z} \\ k \neq 0}} \kappa^{(\ell)} \left(\frac{N}{B}\left(t+k\right)\right) \right\vert.
$$
We conclude on the desired statement by setting the limit $N \to \infty$ and using Assumption (H4). \qed

\section{Proof of Theorem \ref{thm:multipleSources}}\label{sec:proofOfMultipleSources}

Similarly to the proof given in Section \ref{sec:proofs}, we leverage the invariance of the non-degenerate source condition (proved in  Lemma~\ref{lem:invarianceOfNonDegenerateSourceCondition}) defined in \cite{duval_exact_2015} and may restrict, without loss of generality, the analysis to measures $\mu_\star$ with support $\mathcal{T} = \{t_1,\dots,t_S \}$ of the form
\begin{equation}\label{eq:multiSource-reducedMeasure}
	\mu_{\star} = e^{-i\frac{\theta}{2}}\delta\left(t - \Delta/2 \right) + e^{i\frac{\theta}{2}}\delta\left(t+  \Delta/2 \right) + \sum_{s\geq 3} c_s \delta(t-t_s),
\end{equation}
where we impose $t_1 = \Delta/2$ and $t_2 = -\Delta/2$ for some $\Delta > 0$, and $c_1 =  e^{-i\frac{\theta}{2}}$ and $c_2 =  e^{i\frac{\theta}{2}}$ for some $\theta\in[0,\pi]$. The minimal norm vanishing derivative polynomial $Q_V$ associated to the measure $\mu_\star$ in \eqref{eq:multiSource-reducedMeasure} is given by $Q_V(t) = \bm{a}(t)^\ast \diag(\overline{\bm{g}}) \bm{p}_V$, where $\bm{p}_V\in\mathbb{C}^N$ is defined as the solution of the quadratic program
\begin{align}
	\label{eq:minimalVanishingPreCertificate-multipleSources}
	\bm{p}_V = & \argmin_{\bm{p}\in\mathbb{C}^{N}}\left\Vert \bm{p}\right\Vert _{2} \nonumber\\
	\mbox{subject\,to\,}\phantom{=} & \bm{a}\left(t_s\right)^{\ast}\diag(\bar{\bm{g}})\bm{p}= \sign(c_s), \nonumber \\
	& \dot{\bm{a}}\left(t_s\right)^{\ast}\diag(\bar{\bm{g}})\bm{p}=0, \qquad s = 1,\dots,S.
\end{align}
Invoking from \cite{duval_exact_2015} a generalization of Lemma \ref{lem:equivalence} to the case of measures with an arbitrary number $S$ of point sources, it is sufficient to show that $Q_V(t)$ verifies the conditions
\begin{subequations}
	\label{eq:multiSource-minPolynomialConstraints}
	\begin{align}
		\left\vert Q_V(t) \right\vert & < 1, \quad \forall t \notin \mathcal{T},\\
		\frac{{\rm d}^{2}}{{\rm d}t^{2}}\left| Q_V \left(t_s\right) \right| & <0, \quad s = 1,\dots,S,
	\end{align}
\end{subequations}
for a large enough $\Delta$ for any $N\geq N_0$ to conclude on the support stability of the Beurling-LASSO estimator for any measure $\mu_{\star}$ satisfying \eqref{eq:multiSources-separationCondition}.

To proceed, we study, as in the proof of Theorem \ref{thm:stableResolutionLimit}, the asymptotic behavior of the polynomial $Q_V(t)$ when $N\to\infty$. Define for any $s = 1,\dots,S$, the subspace $E_s \subset \mathbb{C}^N$ as
\begin{align}
	E_s = \spann \left\{ \diag(\bm{g}) \bm{a}(t_s), \diag(\bm{g}) \dot{\bm{a}}(t_s) \right\},
\end{align}
and denote by $F = E_1 + E_2$ the sum of the two subspaces associated with the two close point sources $t_1$ and $t_2$. It is clear, from an orthogonality projection argument, that  $\bm{p}_V \in F + \sum_{s=3}^S E_s$. Let $\bm{p}_F \in F$ be the solution of the quadratic program \eqref{eq:minimalVanishingPreCertificate} associated with the two close point sources $t_1$ and $t_2$. Similarly, for any $s\geq 3$ denote by $\bm{p}_{E_s} \in E_s$ the unique solution of the quadratic program
\begin{align}
	\label{eq:minimalVanishingPreCertificateOnSubspaceE}
	\bm{p}_{E_S} = & \argmin_{\bm{p}\in\mathbb{C}^{N}}\left\Vert \bm{p}\right\Vert _{2} \nonumber \\
	\mbox{subject\,to\,}\phantom{=} & \bm{a}\left( t_s \right)^{\ast}\diag(\bar{\bm{g}})\bm{p}= \sign(c_s), \nonumber \\
	& \dot{\bm{a}}\left(t_s\right)^{\ast}\diag(\bar{\bm{g}})\bm{p}=0,
\end{align}
which is equal to
\begin{equation}
	\bm{p}_{E_s} = \frac{\sign(c_s)}{K(0)} \diag(\bm{g}) \bm{a}(t_s)
\end{equation}
for all $s = 3,\dots, S$. We wish to show that $\bm{p}_{V}$ converges to $\bm{p}_{F} + \sum_{s=3}^S \bm{p}_{E_s}$ in the limit of $N$ to get the asymptotic properties of $Q_V(t)$. To that end, we define the matrix $\bm{A}_{E_s}\in\mathbb{C}^{N\times 2}$ for $s=1,\dots,S$ as
\[
	\bm{A}_{E_s} = \left[\diag(\bm{g}) \bm{a}(t_s)/\sqrt{K(0)}, \diag(\bm{g}) \dot{\bm{a}}(t_s)/ \sqrt{\left\vert K^{\pprime}(0) \right\vert}  \right],\quad s = 1,\dots,S.
\]
It is immediate that the columns of $\bm{A}_{E_s}$ form an orthonormal basis of $E_s$. Let  $\bm{A}_{F} \in \mathbb{C}^{N\times 4}$ and $\bm{A} \in \mathbb{C}^{N \times 2S}$ be the concatenations
\begin{align*}
	\bm{A}_{F} &= [\bm{A}_{E_1}, \bm{A}_{E_2}], \\
	\bm{A} &= [\bm{A}_{F}, \bm{A}_{E_3},\dots, \bm{A}_{E_s}] =  [\bm{A}_{E_1}, \bm{A}_{E_2}, \bm{A}_{E_3},\dots, \bm{A}_{E_s}],
\end{align*}
and $\bm{w} = [\sign(c_1),0,\sign(c_2),0,\dots \sign(c_s),0]^{\top}\in\mathbb{C}^{2S}$. By the orthogonal projection theorem, we have that
\begin{subequations}\label{eq:expression-partialProjections}
\begin{align}
 \bm{p}_F &= K(0)^{-\frac{1}{2}} \bm{A}_F \left( \bm{A}_F^\ast \bm{A}_F \right)^{-1} [\sign(c_1),0,\sign(c_2),0]^\top,\\
 \bm{p}_{E_s} &= K(0)^{-\frac{1}{2}} \bm{A}_{E_s} \left( \bm{A}_{E_s}^\ast \bm{A}_{E_s} \right)^{-1} [\sign(c_s),0]^\top, \quad s = 3,\dots S,\\
 \bm{p}_V &= K(0)^{-\frac{1}{2}} \bm{A} \left( \bm{A}^\ast \bm{A} \right)^{-1} \bm{w}.
\end{align}
\end{subequations}
Moreover, denote by $\bm{G} \in \mathbb{C}^{2s \times 2s}$ the matrix
\[
	\bm{G} =
\begin{bmatrix}\bm{A}_{F}^\ast \bm{A}_{F} & \bm{0} & \cdots & \bm{0}\\
\bm{0} & \bm{A}_{E_{3}}^\ast \bm{A}_{E_{3}} & \cdots & \bm{0}\\
\vdots & \vdots & \ddots & \vdots\\
\bm{0} & \bm{0} & \cdots & \bm{A}_{E_{s}}^\ast \bm{A}_{E_{s}}
\end{bmatrix},
\]
then \eqref{eq:expression-partialProjections} implies that
\begin{equation}\label{eq:difference-vanishing-Projected}
\bm{p}_V - \bm{p}_{F} - \sum_{s=3}^S \bm{p}_{E_s} = K(0)^{-1/2} \bm{A} \left( \left( \bm{A}^\ast \bm{A} \right)^{-1}  - \bm{G}^{-1}
\right)\bm{w}.
\end{equation}

Next, we show that the difference $\left( \bm{A}^\ast \bm{A} \right)^{-1}  - \bm{G}^{-1}$ is small in norm when $N\to \infty$. To proceed, we start by noticing that
\[
	\bm{A}^\ast\bm{A} - \bm{G} =
\begin{bmatrix} \bm{0} & \bm{A}_{F}^\ast \bm{A}_{E_3}& \cdots & \bm{A}_{F}^\ast \bm{A}_{E_s}\\
\bm{A}_{E_3}^\ast \bm{A}_{F} & \bm{0} & \cdots & \bm{A}_{E_3}^\ast \bm{A}_{E_s}\\
\vdots & \vdots & \ddots & \vdots\\
\bm{A}_{E_s}^\ast \bm{A}_{F} & \bm{A}_{E_s}^\ast \bm{A}_{E_3} & \cdots & \bm{0}
\end{bmatrix}.
\]
Denote by $\left\Vert \bm{M} \right\Vert_{\mathrm{max}} = \max_{i,j} \left\vert m_{i,j} \right\vert$ the max-norm of $\bm{M}$, which is the maximum of the absolute values of the entries $m_{i,j}$ of $\bm{M}$. A direct calculation establishes that each off-diagonal block of $\bm{A}^\ast\bm{A} - \bm{G}$ verifies
\begin{subequations}\label{eq:limitsOfBlocks}
	\begin{align}
		\left\Vert \bm{A}_{F}^\ast  \bm{A}_{E_s} \right\Vert_{\mathrm{max}} &= \max \left\{
			\left\vert \frac{K(t_s \pm \Delta/2)}{K(0)} \right\vert ,
			\frac{\left\vert K^{\prime}(t_s \pm \Delta/2)\right\vert}{\sqrt{\left\vert K(0)K^{\pprime}(0) \right\vert }} ,
			\left\vert \frac{K^{\pprime}(t_s \pm \Delta/2)}{K^{\pprime}(0)} \right\vert
			\right\}, \label{eq:diagBlockExpression}\\
			\left\Vert \bm{A}_{E_{s^\prime}}^\ast  \bm{A}_{E_s} \right\Vert_{\mathrm{max}} &= \max \left\{
			\left\vert \frac{K(t_s -t_{s^\prime})}{K(0)} \right\vert ,
			\frac{\left\vert K^{\prime}(t_s -t_{s^\prime})\right\vert}{\sqrt{\left\vert K(0)K^{\pprime}(0) \right\vert }} ,
			\left\vert \frac{K^{\pprime}(t_s -t_{s^\prime})}{K^{\pprime}(0)} \right\vert
			\right\}, \quad \forall s,s^\prime \geq 3 \mbox{ and } s\neq s^\prime. \label{eq:diagBlockExpression-2}
	\end{align}
\end{subequations}
Taking the limit $N\to \infty$ in \eqref{eq:diagBlockExpression} yields
\begin{align}
	\lim_{N\to\infty} \left\Vert \bm{A}_{F}^\ast  \bm{A}_{E_s} \right\Vert_{\mathrm{max}} &= \lim_{N\to \infty} \max \left\{
		 \left\vert  \frac{\kappa(N t_s / B \pm \gamma/2)}{\kappa(0)} \right\vert ,
		\frac{\left\vert \kappa^{\prime}(N t_s / B \pm \gamma/2) \right\vert}{\sqrt{\left\vert \kappa(0)\kappa^{\pprime}(0) \right\vert }} ,
		\left\vert \frac{\kappa^{\pprime}(N t_s / B \pm \gamma/2)}{\kappa^{\pprime}(0)} \right\vert
		\right\} \nonumber\\
		&= \max \left\{
			\lim_{ \left\vert \tau \right\vert \to \infty} \left\vert  \frac{\kappa(\tau)}{\kappa(0)} \right\vert,
		\lim_{ \left\vert \tau \right\vert \to \infty} \left\vert  \frac{\kappa^\prime(\tau)}{\sqrt{\left\vert \kappa(0)\kappa^{\pprime}(0) \right\vert }} \right\vert,
		\lim_{ \left\vert \tau \right\vert \to \infty} \left\vert  \frac{\kappa^{\pprime}(\tau)}{\kappa^{\pprime}(0)} \right\vert
		\right\} \nonumber\\
		&= 0,\nonumber
	\end{align}
	where the first equality comes from Lemma \ref{lem:convergenceOfK}, the second equality from the assumption \eqref{eq:multiSources-separationCondition} which implies $\left\vert N t_s / B \pm \gamma/2 \right\vert \to \infty$, $s=3,\dots,S$, when $N\to\infty$, and the last equality holds under assumptions (H2) and (H3) by the application of the Riemann-Lebesgue lemma on the integrable functions $\mathcal{F}\left\{ \kappa^{(\ell)}\right\}(f) = (i2\pi f)^{\ell} \left\vert G(f) \right\vert^2$, $\ell=0,1,2$. An analogous reasoning on \eqref{eq:diagBlockExpression-2} leads to
\(		\lim_{N\to\infty} \left\Vert \bm{A}_{E_{s^\prime}}^\ast  \bm{A}_{E_s} \right\Vert_{\mathrm{max}} = 0,
\)
for all $s,s^\prime \geq 3$ and $s\neq s^\prime$. We conclude that
\begin{equation}\label{eq:convergenceOfBlockMatrix}
	\lim_{N\to\infty} \left\Vert \bm{A}^\ast \bm{A} - \bm{G} \right\Vert_{\mathrm{max}} = 0.
\end{equation}
Next, by Lemma \ref{lem:convergenceOfK}, the matrix $\bm{G}$ admits an invertible limit, and we denote $\overline{\bm{G}} = \lim_{N \to \infty} \bm{G}$. By continuity of the inversion $\bm{M} \mapsto \bm{M}^{-1}$ at $\overline{\bm{G}}$, we have $\lim_{N\to\infty} \bm{G}^{-1} = \overline{\bm{G}}^{-1}$. Moreover, from \eqref{eq:convergenceOfBlockMatrix} we have
\[
	\lim_{N\to\infty} \left\Vert \bm{A}^\ast \bm{A} - \overline{\bm{G}} \right\Vert_{\mathrm{max}} \leq \lim_{N\to\infty}   \left( \left\Vert \bm{A}^\ast \bm{A} - \bm{G} \right\Vert_{\mathrm{max}} +  \left\Vert \bm{G} - \overline{\bm{G}} \right\Vert_{\mathrm{max}} \right) = 0,
\]
implying by continuity that
 $\lim_{N\to\infty} \left(\bm{A}^\ast\bm{A}\right)^{-1} = \overline{\bm{G}}^{-1}$.
Let by $\left\Vert \cdot \right\Vert_2$ and $\left\Vert \cdot \right\Vert_{\mathrm{F}}$ the spectral norm and the Frobenius norm of a matrix, respectively. We have by the previous arguments that
\[
	\lim_{N\to\infty} \left\Vert \left(\bm{A}^\ast \bm{A}\right) ^{-1} - \bm{G}^{-1} \right\Vert_2 \leq \lim_{M\to\infty} \left( \left\Vert \left(\bm{A}^\ast\bm{A}\right) ^{-1} - \overline{\bm{G}}^{-1} \right\Vert_2 + \left\Vert \overline{\bm{G}}^{-1}  - \bm{G}^{-1} \right\Vert_2 \right) = 0,
\]
since $S$ is finite and does not grow with $N$. This, together with \eqref{eq:difference-vanishing-Projected}, yields
\begin{align}
	\left\Vert \bm{p}_V - \bm{p}_{F} - \sum_{s=3}^S \bm{p}_{E_s} \right\Vert_2 &=  K(0)^{-1/2} \left\Vert \bm{A} \left( \left( \bm{A}^\ast \bm{A} \right)^{-1}  - \bm{G}^{-1}
	\right)\bm{w} \right\Vert_2 \nonumber \\
	&\leq  K(0)^{-1/2} \left\Vert \bm{A} \right\Vert_\mathrm{F} \left\Vert \left( \bm{A}^\ast \bm{A} \right)^{-1}  - \bm{G}^{-1} \right\Vert_2 \left\Vert \bm{w} \right\Vert_2 \nonumber \\
	&= \sqrt{2} S  K(0)^{-1/2}  \left\Vert \left( \bm{A}^\ast \bm{A} \right)^{-1}  - \bm{G}^{-1} \right\Vert_2 \nonumber \\
	&\leq \sqrt{2} S  \left( \frac{N \kappa(0)}{B} + o(N) \right)^{-1/2} \left\Vert \left( \bm{A}^\ast \bm{A} \right)^{-1}  - \bm{G}^{-1} \right\Vert_2, \nonumber \\
	&\leq o(N^{-1/2}),\quad N\to\infty,
\end{align}
where the third line follows from $\left\Vert \bm{w} \right\Vert_2 = \sqrt{S}$ and $\left\Vert \bm{A} \right\Vert_{\mathrm{F}} = \sqrt{2S}$, and the fourth line follows from Lemma~\ref{lem:convergenceOfK}.

Denote by $Q_F(t)$ and $Q_{E_s}(t)$ for any $s \geq 3$ the trigonometric polynomials
	\begin{align*}
		Q_F(t) &= \bm{a}(t)^\ast \diag(\overline{\bm{g}}) \bm{p}_F,\\
		Q_{E_s}(t) &= \bm{a}(t)^\ast\diag(\overline{\bm{g}}) \bm{p}_{E_s} = \frac{\sign(c_s)}{K(0)} K(t - t_s),
	\end{align*}
for any $t\in\mathbb{T}$ and $s = 3,\dots,S$. We are now ready to derive the asymptotic behavior of $Q_V(t)$. First of all, for $\ell = 0,1,2$, we have that
\begin{align}\label{eq:multiSource-convergenceOfQv}
	\sup_{t\in\mathbb{T}} \left\vert Q_V^{(\ell)}(t) - Q_F^{(\ell)}(t) - \sum_{s=3}^S Q_{E_s}^{(\ell)}(t) \right\vert &= \sup_{t\in\mathbb{T}} \left\vert  \bm{a}^{(\ell)}(t)^\ast \diag(\overline{\bm{g}}) \left( \bm{p}_V - \bm{p}_F - \sum_{s=3}^S \bm{p}_{E_s} \right) \right\vert \nonumber \\
	&\leq \sup_{t\in\mathbb{T}} \left\Vert \diag(\bm{g}) \bm{a}^{(\ell)}(t)^\ast  \right\Vert_2  \left\Vert \bm{p}_V - \bm{p}_F - \sum_{s=3}^S \bm{p}_{E_s} \right\Vert_2 \nonumber \\
	&\leq \left\Vert \bm{g} \right\Vert_\infty \sup_{t\in\mathbb{T}} \left\Vert\bm{a}^{(\ell)}(t)  \right\Vert_2  \left\Vert \bm{p}_V - \bm{p}_F - \sum_{s=3}^S \bm{p}_{E_s} \right\Vert_2 \nonumber \\
	&= \left\Vert \bm{g} \right\Vert_\infty (2\pi)^\ell \sqrt{\sum_{k=-n}^n k^{2\ell}}  \left\Vert \bm{p}_V - \bm{p}_F - \sum_{s=3}^S \bm{p}_{E_s} \right\Vert_2 \nonumber \\
	&\leq \left\Vert \bm{g} \right\Vert_\infty \mathcal{O} \left( N^{\ell + \frac{1}{2}} \right) o \left( N^{-\frac{1}{2}} \right) \nonumber \\
	&\leq o \left(N^\ell\right),\qquad N \to \infty,
\end{align}
where we used in the last inequality assumption (H3) to bound
\[
	\left\Vert \bm{g} \right\Vert_\infty = \max \left\{ \left\vert G\left(\frac{kB}{N}\right) \right\vert \right\}_{k=-n}^n \leq \sup_{f\in[-\frac{B}{2},\frac{B}{2}]} \left\vert G(f) \right\vert < \infty.
\]

Next, we denote by $\mathcal{Q}_V(\tau)$ the function
\[
	\mathcal{Q}_V(\tau) = \mathcal{Q}_F(\tau) + \sum_{s=3}^S \mathcal{Q}_{E_s} (\tau), \quad \forall \tau \in \mathbb{R},
\]
where $\mathcal{Q}_F(\tau)$ is the function given in \eqref{eq:qVExpression} and $\mathcal{Q}_{E_s}(\tau) = \frac{\sign(c_k)}{\kappa(0)} \kappa(\tau - \frac{N t_s}{B})$ for any $s\geq 3$. It follows that
\begin{align}\label{eq:multiSource-convergenceOfQV}
&\quad \sup_{t\in\mathbb{T}} \left\vert \left(\frac{B}{N}\right)^\ell Q_V^{(\ell)}(t) - \mathcal{Q}_V^{(\ell)}\left(\frac{Nt}{B}\right)\right\vert \nonumber\\
&=	\sup_{t\in\mathbb{T}} \left\vert \left(\frac{B}{N}\right)^\ell Q_V^{(\ell)}(t) - \mathcal{Q}_F^{(\ell)}\left(\frac{Nt}{B}\right) - \sum_{s=3}^S \mathcal{Q}_{E_s}^{(\ell)} \left(\frac{Nt}{B}\right) \right\vert \nonumber\\
	&\leq \left(\frac{B}{N} \right)^{\ell} \sup_{t\in\mathbb{T}} \left\vert Q_V^{(\ell)}(t) - Q_F^{(\ell)}(t) - \sum_{s=3}^S Q_{E_s}^{(\ell)}(t) \right\vert  + \sup_{t\in\mathbb{T}} \left\vert \left(\frac{B}{N}\right)^\ell Q_F^{(\ell)}(t)  - \mathcal{Q}_F^{(\ell)}\left(\frac{Nt}{B}\right) \right\vert \nonumber \\
	&\qquad + \sum_{s=3}^S \sup_{t\in\mathbb{T}} \left\vert \left(\frac{B}{N}\right)^\ell Q_{E_s}^{(\ell)}(t)  - \mathcal{Q}_{E_s}^{(\ell)}\left(\frac{Nt}{B}\right) \right\vert \nonumber \\
	&\leq o(1), \qquad N\to \infty,
\end{align}
where the first inequality follows from the triangle inequality, and the second inequality follows from the fact that each term in the line above is controlled by \eqref{eq:multiSource-convergenceOfQv} and \eqref{eq:uniformConvergeceOfQ} and Lemma \ref{lem:convergenceOfK}, respectively.

With \eqref{eq:multiSource-convergenceOfQV} in place, similarly to the proof presented in Section \ref{sec:proofs}, we argue that the non-degenerate source condition is verified for any $N \geq N_0$  for $N_0$ sufficiently large if the limit function $\mathcal{Q}_V(\tau)$ meets the conditions
\begin{subequations}
	\label{eq:conditionOnLimitQ}
			\begin{align}
			\label{eq:multiSources-conditonOnLimitQ-concavity}
			\lim_{N\to \infty} \frac{{\rm d}^{2}\left|{\mathcal{Q}_{V}}\right|}{{\rm d}\tau^{2}}\left(\frac{N t_s}{B}\right) & < 0, \quad s =1,\dots, S, \\
			\label{eq:multiSouces-conditonOnLimitQ-modulus}
			\lim_{N\to \infty}\left\vert \mathcal{Q}_{V}\left(\frac{N t}{B}\right) \right\vert &< 1, \quad \forall t \notin \left\{ t_s \right\}.
			\end{align}
	\end{subequations}
	In order to verify \eqref{eq:conditionOnLimitQ}, we start by noticing that for any $\tau \in\mathbb{R}$ and $\ell = 0,1,2$,
	\begin{align}\label{eq:boundOnQv-multiSpikes}
		\lim_{N\to \infty}\left\vert \mathcal{Q}^{(\ell)}_{V}\left(\tau \right) \right\vert &\leq \lim_{N \to \infty} \left\vert \mathcal{Q}^{(\ell)}_{F}\left(\tau \right) \right\vert + \sum_{s=3}^S \lim_{N \to \infty} \left\vert {\mathcal{Q}}_{E_s}^{(\ell)}\left(\tau \right) \right\vert.
	\end{align}

	The condition \eqref{eq:multiSources-conditonOnLimitQ-concavity} can be verified by picking $t_{s_\star} \in \{ t_1, \dots ,t_S \}$ and distinguishing two cases:
\begin{itemize}
	\item If $s_\star = 1,2$, and we have that for $\ell = 0,1,2$ that
		\begin{align}
	\lim_{N \to \infty} \left\vert {\mathcal{Q}}^{(\ell)}_{E_s}\left(\frac{N t_{s_{\star}} }{B}\right) \right\vert &= \lim_{N\to\infty} \left\vert \frac{\kappa^{(\ell)} \left( \frac{N}{B}(t_{s_\star}-t_s)\right)}{\kappa(0)} \right\vert = \lim_{\left\vert \tau \right\vert \to \infty}\left\vert \frac{\kappa^{(\ell)}(\tau)}{\kappa(0)} \right\vert = 0, \quad \forall s = 3,\dots,S \label{eq:multiSpikes-case2.1}
\end{align}
where we used the separation assumption and the Riemann-Lebesgue on the integrable functions $\mathcal{F}^{(\ell)}(\kappa)$, $\ell = 0,1,2$, in \eqref{eq:multiSpikes-case2.1}. This leads to
\[
	\lim_{N\to \infty} \frac{{\rm d}^{2}\left|{\mathcal{Q}_{V}}\right|}{{\rm d}\tau^{2}}\left(\frac{N t_{s_\star}}{B}\right) =  \frac{{\rm d}^{2}\left|{\mathcal{Q}_{F}}\right|}{{\rm d}\tau^{2}}\left(\frac{N t_{s_\star}}{B}\right) < 0
\]
by \eqref{eq:boundOnQv-multiSpikes}, and by going to the limit $N\to\infty$ in the expression \eqref{eq:modulusDerivativeExpression}, and invoking the derivations on $\frac{{\rm d}^{2}\left|{\mathcal{Q}_{V}}\right|}{{\rm d}\tau^{2}}$ presented in Section~\ref{subsec:Minimal-vanishing-pre-certificate} under the separation condition \eqref{eq:multiSources-separationCondition}.

	\item If $s_\star = 3,\dots,S$, a similar reasoning yields that for $\ell = 0,1,2$
	\begin{align*}
		\lim_{N \to \infty} \left\vert {\mathcal{Q}}^{(\ell)}_{E_s}\left(\frac{N t_{s_{\star}} }{B}\right) \right\vert &= \lim_{N\to\infty} \left\vert \frac{\kappa^{(\ell)} \left( \frac{N}{B}(t_{s_\star}-t_s)\right)}{\kappa(0)} \right\vert = \lim_{\left\vert \tau \right\vert \to \infty}\left\vert \frac{\kappa^{(\ell)}(\tau)}{\kappa(0)} \right\vert = 0, \quad \forall s \in \left\{ 3,\dots, S\right\} \backslash \{ s_\star \},
		 \\
		\lim_{N \to \infty} \left\vert {\mathcal{Q}}^{(\ell)}_{F}\left(\frac{N t_{s_{\star}} }{B}\right) \right\vert &= \lim_{\left\vert \tau \right\vert \to \infty}\left\vert \mathcal{Q}^{(\ell)}_{F}\left(\tau \right) \right\vert = 0.
	\end{align*}
	We analogously  obtain that
	\[
	\lim_{N\to \infty} \frac{{\rm d}^{2}\left|{\mathcal{Q}_{V}}\right|}{{\rm d}\tau^{2}}\left(\frac{N t_{s_\star}}{B}\right) =  \frac{{\rm d}^{2}\left|{\mathcal{Q}_{E_{s_\star}}}\right|}{{\rm d}\tau^{2}}\left(\frac{N t_{s_\star}}{B}\right) = \frac{\kappa^{\pprime} (0)}{\kappa(0)}< 0.
\]
\end{itemize}

Finally, to show that \eqref{eq:multiSouces-conditonOnLimitQ-modulus} holds, for any $t \in \mathbb{T} \backslash \left\{ t_s \right\}$, we denote by $s_\star$ the index of the closest point source in the sense of the wrap-around distance
\[
	t_{s_\star} = \min_{s=1,\dots, S} | t - t_s |_{\mathbb{T}}.
\]
Here again, we distinguish two cases:
\begin{itemize}
	\item If $s_\star = 1,2$, we have that
	\begin{subequations}
			\begin{align}
		\lim_{N \to \infty} \left\vert {\mathcal{Q}}_{E_s}\left(\frac{N t }{B}\right) \right\vert &= \lim_{N\to\infty} \left\vert \frac{\kappa \left( \frac{N}{B}(t-t_s)\right)}{\kappa(0)} \right\vert = \lim_{\left\vert \tau \right\vert \to \infty}\left\vert \frac{\kappa(\tau)}{\kappa(0)} \right\vert = 0, \quad \forall s = 3,\dots,S, \label{eq:multiSpikes-case1.1}\\
		\lim_{N\to \infty}\left\vert \mathcal{Q}_{F}\left(\frac{N t}{B}\right) \right\vert &< 1 ,\label{eq:multiSpikes-case1.2}
	\end{align}
\end{subequations}
	where we used the Riemann-Lebesgue on $\mathcal{F}(\kappa)$ in \eqref{eq:multiSpikes-case1.1}, and by invoking the derivations on $\mathcal{Q}_F$ of Section \ref{subsec:Minimal-vanishing-pre-certificate} under the separation condition \eqref{eq:multiSources-separationCondition} in \eqref{eq:multiSpikes-case1.2}.
	\item If $s_\star = 3,\dots S$, we have that
	\begin{subequations}
	\begin{align}
		\lim_{N \to \infty} \left\vert {\mathcal{Q}}_{E_s}\left(\frac{N t }{B}\right) \right\vert &= \lim_{N\to\infty} \left\vert \frac{\kappa \left( \frac{N}{B}(t-t_s)\right)}{\kappa(0)} \right\vert = \lim_{\left\vert \tau \right\vert \to \infty}\left\vert \frac{\kappa(\tau)}{\kappa(0)} \right\vert = 0,\quad \forall s \in \left\{ 3,\dots, S \right\} \backslash \{s_\star\} ,\\
		\lim_{N \to \infty} \left\vert {\mathcal{Q}}_{E_{s_\star}}\left(\frac{N t }{B}\right) \right\vert &= \lim_{N\to\infty} \left\vert \frac{\kappa \left( \frac{N}{B}(t-t_{s_\star})\right)}{\kappa(0)} \right\vert < 1,\\
		\lim_{N\to \infty}\left\vert \mathcal{Q}_{F}\left(\frac{N t}{B}\right) \right\vert &= \lim_{\left\vert \tau \right\vert \to \infty}\left\vert \mathcal{Q}_{F}\left(\tau \right) \right\vert = 0.
	\end{align}
\end{subequations}
\end{itemize}
Summing all the terms in both cases, and invoking \eqref{eq:boundOnQv-multiSpikes} concludes on \eqref{eq:multiSouces-conditonOnLimitQ-modulus}.


In conclusion, under the conditions of Theorem \ref{thm:multipleSources}, there must exists $N_0\in\mathbb{N}$ such that the polynomial $\mathcal{Q}_V(\tau)$ verifies the conditions \eqref{eq:multiSource-minPolynomialConstraints}. We conclude that the measure $\mu_\star$ verifies the non-degenerate source condition, implying the support stability of the Beurling-LASSO estimator. \qed

\bibliography{Atomic.bib}
\bibliographystyle{IEEEtran}


\end{document}